\documentclass[11pt]{article}
\usepackage{amssymb}

\usepackage{graphicx}
\usepackage{amsmath}

\input{tcilatex}

\begin{document}

\vspace*{-.6in} \thispagestyle{empty}
\begin{flushright}
ITFA--2003--05\\
CFP--2003--01\\
\end{flushright}
\vspace{.2in} {\Large
\begin{center}
{\bf On the Classical Stability of Orientifold Cosmologies}
\end{center}}
\vspace{.2in}
\begin{center}
\textbf{Lorenzo Cornalba}$^{\dagger}$\footnote{lcornalb@science.uva.nl}\ \ 
and\ \ \textbf{Miguel S. Costa}$^{\ddagger}$\footnote{miguelc@fc.up.pt}
\\
\vspace{.2in} $^{\dagger}$\emph{Instituut voor Theoretische Fysica,
Universiteit van Amsterdam\\Valckenierstraat 65, 1018 XE Amsterdam, The Netherlands}\\
\vspace{.1in}
$^{\ddagger}$\emph{Centro de  F\'\i sica do Porto, Departamento de F\'\i sica\\ 
Faculdade de Ci\^encias, Universidade do Porto\\
Rua do Campo Alegre, 687 - 4169-007 Porto, Portugal
}
\end{center}

\vspace{.3in}

\begin{abstract}

\textit{We analyze the classical stability of string cosmologies driven by the dynamics of
orientifold planes. These models are related to time--dependent orbifolds, and resolve
the orbifold singularities which are otherwise problematic by introducing orientifold planes.
In particular, we show that the instability discussed by Horowitz and Polchinski
for pure orbifold models is resolved by the presence of
the orientifolds. Moreover, we discuss the issue of stability of the
cosmological Cauchy horizon, and we show that it is stable to small perturbations
due to in--falling matter.}

\end{abstract}
\newpage

\section{Introduction}

Over the past year there has been a renewed interest in the study of time--dependent string
backgrounds [1--33]. The main objective of this programme is to learn about the origin of the
(possibly apparent) cosmological singularity using the powerful techniques of string theory. The logic
behind this investigation was to start with an exact conformal field theory, \textit{e.g.} strings in
flat space, and to consider an orbifold of the theory with a time--dependent quotient space \cite{HS}. 
These simple orbifold constructions were, however, quite generally shown to be unstable by
Horowitz and Polchinski \cite{HP} 
\footnote{The exception is the null--brane \cite{SimonFigueroa} with extra non--compact directions \cite{LMS2}, 
which is an exact and regular bounce in string theory free of singularities. Notice, however, 
that this space is not a cosmology since it is supersymmetric and depends on a null, 
not a time--like, direction.}. These instabilities are 
associated to the formation of large black holes whenever a particle is coupled to the geometry
and are non-perturbative in the string coupling. For this reason a perturbative string resolution of 
the time--dependent orbifold singularities seems inappropriate. 

In \cite{CC} we investigated the orbifold of flat space--time by a boost and translation
transformation. This orbifold suffers from the Horowitz--Polchinski instability. However, 
an aditional essential  ingredient of the model was introduced in \cite{CCK}:
\textit{the time--like orbifold singularity is an orientifold, i.e. a brane with negative tension}.
One of the results that we shall show is that the orientifold provides precisely the non--perturbative
string resolution of the singularity, modifying the gravitational interaction in its proximity.
In fact, after briefly revising the Horowitz--Polchinski argument and the orientifold cosmology 
construction, we shall show in Section 4, with a tractable example in dimension $3$,
that \textit{due to the presence of the orientifolds large black holes are not formed.}

In any pre big--bang scenario, there is generically a possible classic instability associated
with the propagation of matter through the bounce, which is analogous to the propagation of matter 
between the outer and inner horizons of charged black--holes \cite{PenroseSimpson,CH}.
Let us consider a perturbation at some finite time in the far past, when the universe is contracting.
In the case of orientifold cosmology 
there is a future cosmological horizon where the contracting phase ends 
(see also \cite{KounnasLust,Q0,Q1}), therefore one needs to worry about such perturbations 
diverging at the horizon, thus creating a big crunch space--time.
We shall show in Section 5 that such perturbations remain finite (of course, they can grow and 
create small black holes, but this does not change the causal cosmological structure of the geometry). 
Furthermore, we shall see that the fluctuations that destabilize the geometry are precisely those 
that already destabilize the Minkowski vacuum.

\section{The Horowitz--Polchinski Problem\label{HPproblem}}

In \cite{HP}, Horowitz and Polchinski argue that a large class of time
dependent orbifolds are unstable to small perturbations, due to large
backreactions of the geometry. Their results, which overlap with the string
theory scattering computations of \cite{LMS1,LMS2}, do not rely on string
theory arguments, and are obtained simply within the framework of classical
general relativity.

The argument of Horowitz and Polchinski is quite simple and runs as follows.
They analyze the orbifold of $D$--dimensional Minkowski space $\mathbb{M}^{D}
$ by the action of a discrete isometry $e^{\kappa }$ where $\kappa $ is a
Killing vector, and consider, in the quotient orbifold space $\mathbb{M}%
^{D}/e^{\kappa }$, a small perturbation. This perturbation naturally
corresponds to an infinite sequence of perturbations in the covering space $%
\mathbb{M}^{D}$, related to each other by the action of $e^{\kappa }$. They
consider, quite generally, a light ray with world--line $\Omega \subset 
\mathbb{M}^{3}$, together with all of its images 
\begin{equation*}
\Omega _{n}=\,e^{n\kappa }\Omega \,\ \ \ \ \ \ \ \ \ \ \ \ \ \ \ \ \ \ \ \ \
\ \ \ \ \left( n\in \mathbb{Z}\right) 
\end{equation*}
for any integer $n$. The basic claim of \cite{HP} is that, for a wide range
of time dependent orbifolds, the backreaction of gravity to the presence
of this infinite sequence of light rays $\Omega _{n}$ will inevitably
produce large black holes. As we already mentioned, this statement is a
claim in pure classical gravity theory and to prove it Horowitz and
Polchinski consider the scattering of two light rays $\Omega _{0}$ and $%
\Omega _{n}$. They compute, given $\kappa $, the impact parameter $b_{n}$
and the center of mass energy $\mathcal{E}_{n}$ of the scattering, and show
that, for generic time dependent orbifolds, one has that $G\mathcal{E}%
_{n}\gg b_{n}^{D-3}$ for $n$ sufficiently large. The two image particles
will be well within the Schwarzchild radius corresponding to the center of
mass energy, and will form a black hole.

It is quite clear that, if the above argument is correct, it must hold also
for a continuous distribution of light rays 
\begin{equation}
e^{t\kappa }\Omega \,\ \ \ \ \ \ \ \ \ \ \ \ \ \ \ \ \ \ \ \ \ \ \ \ \
\left( t\in \mathbb{R}\right)   \label{HPsurface}
\end{equation}
since the problematic interactions occur between rays separated by $\Delta
t\gg 1$, and therefore should be completely insensitive to the distribution
of the essentially parallel light rays when the parameter $t$ varies by $%
\Delta t\sim 1$. For continuous values of $t$, equation ($\ref{HPsurface}$)
defines a surface $S$ in $\mathbb{M}^{D}$, which we call the
Horowitz--Polchinski (HP) surface. Again, the backreaction of gravity to the
matter distribution on $S$ should form black holes by the previous argument.
On the other hand, the continuous distribution is easier to treat
analytically since it is invariant under the action of $\kappa $. Therefore,
the full solution with the gravitational backreaction will also have $\kappa 
$ as a Killing vector. This fact simplifies explicit computations and
therefore, from now on, we will only consider the continuous case.

In the rest of the paper we will be interested exclusively in a specific
choice of orbifold, which describes, when correctly interpreted in $M$%
--theory, the dynamics of an orientifold/anti--orientifold pair. We postpone
the discussion of the correct $M$--theory embedding and the crucial distinction
between the orbifold and the orientifold interpretation to section 
\ref{orientifold}, and we describe now the orbifold along the lines of \cite
{HP}.

We write $\mathbb{M}^{D}=\mathbb{M}^{3}\times \mathbb{E}^{D-3}$ and we
parametrize $\mathbb{M}^{3}$ with coordinates $X^{\pm }$, $Y$, with line element
\begin{equation}
-dX^{+}dX^{-}+dY^{2}\,.  \label{flat}
\end{equation}
Points in $\mathbb{M}^{D}$ are identified along the orbits of the Killing
vector field $\kappa $ which corresponds to boost in the $X^{\pm }$ plane
and translation in the orthogonal direction $Y$ 
\begin{equation}
\kappa =X^{+}\partial _{+}-X^{-}\partial _{-}+\,\partial _{Y}\,.
\label{killing}
\end{equation}
For this special choice of $\kappa $ one has $b_{n}\sim n$ and $\mathcal{E}%
_{n}\sim e^{n}$ and the bound $G\mathcal{E}_{n}\gg b_{n}^{D-3}$ is always
satisfied for $n$ large \cite{HP}. The directions along $\mathbb{E}^{D-3}$
then play no role (except to determine the strength of the gravitational
interaction) and we will omit any reference to them, working effectively in
three dimensions.

To study the geometry of the HP surface $S$, we parametrize the light ray $%
\Omega $ as 
\begin{equation*}
A^{a}+sV^{a}\text{\thinspace },
\end{equation*}
where $V^{a}$ is a null vector and $s$ is an affine parameter. If we
consider the generic case\footnote{%
When $V^{+}V^{-}=0$ the images $e^{t\kappa }\Omega $ correspond to parallel
light rays, and do not form black holes.} when $V^{+}V^{-}\neq 0$ then, for
some value of $t$, the image ray $e^{t\kappa }\Omega $ will be directed
along the null vector\footnote{%
It could also be directed along $\left( 1,1,-1\right) $. This case is just
the mirror under $X^{\pm }\rightarrow X^{\mp }$ and we do not discuss it.} $%
\left( 1,1,1\right) $. This specific light ray on the surface $S$ will
intersect the plane $X^{+}+X^{-}=0$ at a point with $X^{+}-X^{-}=2a$. By
shifting the variables $s,t,Y$ so that this point corresponds to $s=t=Y=0$
we see that the surface $S$ is given by the following parametrization 
\begin{eqnarray}
X^{\pm } &=&\left( s\pm a\right) e^{\pm t}\, ,  \label{surface} \\
Y &=&\left( s+t\right)\, .  \notag
\end{eqnarray}
\begin{figure}
\begin{center}
\includegraphics{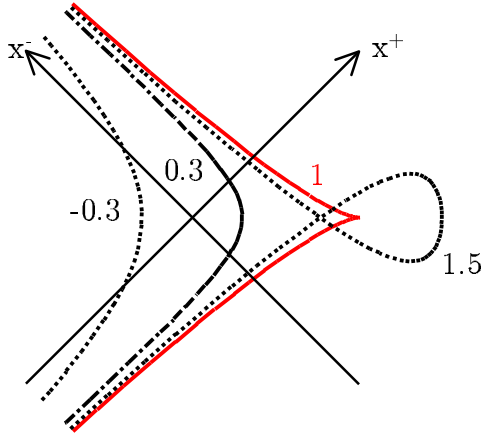}
\end{center}
\caption{The curves given by equation (\ref{surxpxm}),
which represent the surfaces $S_a$ for the values of $a$ 
indicated in the figure. The surface $S_1$ is clearly
singled out, and is the unique null surface.\label{fig7}}
\end{figure}
The only free parameter which labels different surfaces is the constant $a$,
and we will add a label $S_{a}$ to distinguish them. A convenient way to
visualize $S_{a}$ is to move to coordinates 
\begin{eqnarray}
x^{\pm } &=&X^{\pm }e^{\mp Y}\, ,  \label{2dcoord} \\
y &=&Y\, ,  \notag
\end{eqnarray}
so that $\kappa =\frac{\partial }{\partial y}$. Then, since the surface $%
S_{a}$ is invariant under $\kappa $, it is given by a curve in the $%
x^{+},x^{-}$ plane (see Figure \ref{fig7})
\begin{equation}
x^{\pm }=\left( s\pm a\right) e^{\mp s}.  \label{surxpxm}
\end{equation}
This curve can also be viewed as the intersection of the surface $S_{a}$
with the plane $Y=0$ in $\mathbb{M}^{3}$.

Let us now describe some basic properties of the surfaces $S_{a}$. The
induced metric is given by 
\begin{equation*}
2\left( 1-a\right) dsdt+\left( s^{2}-a^{2}+1\right) dt^{2}
\end{equation*}
so it is timelike for all values of $a$ except for $a=1$, when $S$ is a
null surface. For $a\neq 1$ the surfaces are smooth for any value of the
parameters $s,t$. For $a=1$ the surface is, on the other hand, divided into
two smooth parts for $s>0$ and $s<0$ which are joined along the singular
line $s=0$. Moreover, the forward directed light cones, which are tangent to
the null surface $S_{1}$, lie on one side of $S_{1}$ for $s>0$ and jump
discontinuously, along the singular line $s=0$, to the other side of $S_{1}$
for $s<0$.

\subsection{The Three--Dimensional Horowitz--Polchinski Problem\label{3dHP}}

As we described in the previous section, we want to understand, in general,
the gravitational backreaction to a distribution of light--rays in $\mathbb{%
M}^{3}\times \mathbb{E}^{D-3}$ distributed on the surface $S_{a}$
parametrized in $\mathbb{M}^{3}$ by (\ref{surface}) and fixed at a point
along the spectator directions $\mathbb{E}^{D-3}$. For $D\geq 4$ gravity is
non--trivial even in the absence of matter, and the problem is not soluble
exactly. In $D=3$, on the other hand, space is curved only in the presence
of matter and the problem is tractable.

\begin{figure}
\begin{center}
\includegraphics{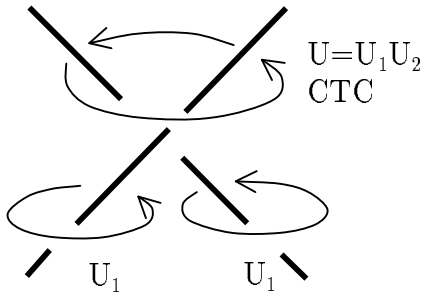}
\end{center}
\caption{Scattering of light rays in three dimensions. The holonomies $U_1$ 
and $U_2$ are null boosts, whereas $U$ is a boost for center of mass energies
greater then a specific threshold. In this case, the geometry has CTC's, as
indicated. \label{fig8}}
\end{figure}

Recall that the basic question is whether black holes are formed as a
consequence of the strong gravitational interactions among the light rays.
On the other hand, black holes in three dimensional flat space do not exist
in the first place, and therefore it seems that the question is not well
posed for $D=3$. This is not correct. In fact, the sign of the formation of
a black hole in three dimensions is not the presence of a singularity but
the existence, in the geometry, of closed timelike curves. To see this,
recall again that, in a two--particle scattering with center of mass energy $%
\mathcal{E}$ and impact parameter $b$, a black hole forms when 
$G\mathcal{E}\gtrsim b^{D-3}$. For $D=3$ the threshold is then 
$G\mathcal{E}\gtrsim 1$ and is independent of the impact
parameter $b$. To understand the meaning of this threshold, recall that the
geometry describing the propagation of a single light ray in 3D is flat
everywhere except along a null line, where curvature is concentrated. The
local geometry is completely characterized by the holonomy in $SO\left(
1,2\right) $ around the light ray, which is given by a specific null boost 
$U_{1}$. In the presence of a second light ray, we have a second holonomy
null boost matrix $U_{2}$, as shown in Figure \ref{fig8}. The combined holonomy $%
U=U_{1}U_{2}$, on the other hand, depends on the center of mass energy of the two
particles. For small center of mass energies $U$ is a rotation. On the other hand, when
the center of mass energy exceeds the threshold value \cite{Gott,Deser,Kabat,Gott2}, $U$
becomes a boost and closed timelike curves are formed which circle both
particles, as shown in Figure \ref{fig8}. Finally, as shown in \cite{Matschull},
one may consider the same dynamics in $AdS_{3}$ space, where black holes
exist \cite{BTZ}. One then finds that, at exactly the same threshold, the
scattering process creates, in the future, a BTZ black hole.

We now see that the three dimensional problem consists in analyzing the
gravitational backreaction to light rays located on the surface (\ref
{surface}), and investigating the existence, in the full geometry, of
closed timelike curves. As we already mentioned, since the surface $S_{a}$
preserves the Killing vector $\kappa $, the full geometry with backreaction
will also have the same Killing vector. We will show two things:

\begin{enumerate}
\item  The geometry will have closed timelike curves. It is important to
stress that we are talking about the \textit{uncompactified} geometry, and
not of the one obtained after identifying by $e^{\kappa }$.

\item  If we excise from the spacetime manifold the region where $\kappa $
is timelike, then the resulting geometry is free from closed timelike
curves. As discussed in \cite{CCK}, the excision is necessary to correctly
embed these geometries in $M$--theory. This procedure introduces a boundary
in the geometry which corresponds, in string/$M$--theory, to the presence of
orientifold planes. We will recall these points more at length in the next
section.
\end{enumerate}

\section{Orientifold versus Orbifold Cosmology\label{orientifold}}

As we just mentioned, and as discussed in detail in \cite{CCK}, when we
embed the orbifold spaces in string/$M$--theory we must excise the regions
with $\kappa ^{2}<0$. The space $\mathbb{M}^{3}$ is naturally divided into
two regions, $\mathcal{X}$ and $\mathcal{Y}$, where the vector $\kappa $ is
spacelike (or null) and timelike, respectively. Since $\kappa ^{2}=1+X^{+}X^{-}$, $%
\mathcal{X}$ and $\mathcal{Y}$ correspond to the regions $X^{+}X^{-}\geq -1$ and 
$X^{+}X^{-}<-1$. We can now consider the quotient manifold 
\begin{equation*}
\mathbb{M}_{Q}^{3}=\mathbb{M}^{3}/e^{\kappa }=\mathcal{X}_{Q}\cup \mathcal{Y}%
_{Q}\, ,
\end{equation*}
where $\mathcal{X}_{Q}=\mathcal{X}/e^{\kappa }$ and $\mathcal{Y}_{Q}=%
\mathcal{Y}/e^{\kappa }$. The action of $\kappa $ is free and therefore the
quotient manifolds are smooth. On the other hand, the full quotient space $%
\mathbb{M}_{Q}^{3}$, although regular and without boundary, has closed
timelike curves which always pass in region $\mathcal{Y}_{Q}$. The space $%
\mathcal{X}_{Q}$, on the other hand, does not have closed timelike curves,
but does have a boundary.

In the usual time dependent orbifold model, one considers the following
compactification of $M$ and $IIA$ theory 
\begin{equation*}
\begin{array}{ccccc}
&  & \mathbb{M}_{Q}^{3}\times \mathbb{T}^{8} &  &  \\ 
& \swarrow _{S^{1}} &  & _{\kappa }\searrow  &  \\ 
\mathbb{M}_{Q}^{3}\times \mathbb{T}^{7} &  & \overset{TST}{\longleftrightarrow }
&  & ?
\end{array}
\end{equation*}
The top line is the $M$--theory vacuum. Going to the left, we are
compactifying along a circle $S^{1}$ of the $8$--torus, and we obtain the
Type IIA orbifold with constant dilaton. Going to the right, we are
compactifying along the Killing vector $\kappa $.
The bottom right corner should be related, as usual,
by a $TST$ duality transformation to the Type IIA\ orbifold on the bottom
left. On the other hand, the IIA supergravity background on the right
is quite peculiar, since it has a complex dilaton field. This is 
because the vector $\kappa $ is timelike in $\mathcal{Y}_{Q}$.

In \cite{CC, CCK} we have analyzed more carefully the bottom right corner of
the diagram and have proposed a related but different compactification of $M$%
--theory, which uses crucially non--perturbative objects of the underlying
string theory. In fact, it was noted in \cite{CC, CCK} that, if one excises
from the $M$--theory vacuum the region $\mathcal{Y}_{Q}$ and dimensionally
reduces the space $\mathcal{X}_{Q}\times \mathbb{T}^{8}$ along $\kappa $,
one obtains a warped IIA supergravity solution with non--trivial dilaton and
RR $1$--form of the form $\mathcal{Z}\times \mathbb{T}^{8}$, where $\mathcal{%
Z}$ is a two--dimensional space with boundary. The boundary of $\mathcal{Z}$
is singular and corresponds to the supergravity fields of an orientifold
plane, which acts as a boundary of spacetime, and has well defined boundary
conditions for the string fields. The geometry is best understood after $T$%
-dualizing along the $8$--torus and describes then the interaction of a $O8$%
--$\overline{O8}$ pair, a system which has also been studied in conformal
field theory in \cite{ADSagnotti,Kachru}. We now have a consistent
compactification scheme given by 
\begin{equation*}
\begin{array}{ccccc}
&  & \widetilde{\mathcal{X}}_{Q}\times \mathbb{T}^{8} &  &  \\ 
& \swarrow _{S^{1}} &  & _{\kappa }\searrow &  \\ 
\widetilde{\mathcal{X}}_{Q}\times \mathbb{T}^{7} &  & \overset{TST}{%
\longleftrightarrow } &  & \widetilde{\mathcal{Z}}\times \mathbb{T}^{8}
\end{array}
\end{equation*}
where we are adding tildes on the spaces to recall that one has to impose
specific boundary conditions on the string fields at the boundaries of these
spaces.

\section{Reducing to Two Dimensions\label{mysect}}

We now move back to the analysis of the three--dimensional
Horowitz--Polchinski problem. As we already mentioned, the full geometry
after backreaction will still have a Killing vector. Therefore, upon
dimensional reduction, the problem is effectively two--dimensional.

Let then the general form of the metric in three dimensions be 
\begin{equation}
ds_{3}^{2}=ds_{2}^{2}+\Phi ^{2}\left( dy+A_{a}dx^{a}\right) ^{2}\,,
\label{compactification}
\end{equation}
where $\frac{\partial }{\partial y}$ is the Killing direction. The 
three--dimensional Hilbert action reduces to 
\begin{equation*}
\int d^{2}x\sqrt{g}\left( \Phi R-\frac{1}{2}\Phi ^{3}F^{2}\right) .
\end{equation*}
The equation of motion for the field strength $F$ is trivial and implies
that the scalar field $\Phi ^{3}\star F$ is constant. By rescaling the
variables $y$, $A_{a}$ and $\Phi ^{-1}$ we can fix the constant to any
desired value (provided it does not vanish) so that 
\begin{equation}
\star F=\frac{2}{\Phi ^{3}}.  \label{norm}
\end{equation}
Then, the equations of motion for the dilaton and the metric can be derived 
from the action
\begin{equation}
\int d^{2}x\sqrt{g}\left( \Phi R-\frac{2}{\Phi ^{3}}\right) .  \label{2Dflat}
\end{equation}
This action is a particular case of two--dimensional dilaton gravity (see 
\cite{Noji,2Dgravity} for reviews and a complete list of references), and we
will find it most convenient to work in this framework from now on. Only at
the end we will connect back to the discussion in three dimensions.

We conclude this section by showing that (\ref{2Dflat}) is also relevant in
the study of the interaction of $O8$--$\overline{O8}$--planes wrapped on a
$8$--torus, which again is naturally a
two--dimensional problem. In general, the system is described
by the following massive IIA background 
\begin{eqnarray}
E^{2}ds_{10}^{2} &=&\Phi \,ds_{2}^{2}+\Phi ^{-1}\,ds^{2}\left( \mathbb{T}%
^{8}\right) \,,  \label{uplift} \\
e^{\phi } &=&g_{s}\Phi ^{-\frac{5}{2}},\,\ \ \ \ \ \ \ \ \ \ \ \ \ \ \star
F_{10}=\frac{2E}{g_{s}}\,.  \notag
\end{eqnarray}
Then the Type IIA action 
\begin{equation*}
\int d^{10}x\sqrt{G_{10}}e^{-2\phi }\left( R_{10}+4\left( \nabla \phi
\right) ^{2}\right) -\frac{1}{2}\int F_{10}\wedge \star F_{10}
\end{equation*}
becomes the simple two--dimensional dilaton--gravity action 
\begin{equation*}
\int d^{2}x\sqrt{g}\left( \Phi R-\frac{2 E^{2}}{\Phi ^{3}}\right)\, ,
\end{equation*}
which again reduces to (\ref{2Dflat}) after rescaling of $\Phi $. Note that,
in conformal field theory, we can construct only a single $O8$--$\overline{O8%
}$ pair with specific tensions and charges. This then implies (see \cite{CCK}
for more details) that 
\begin{equation}
El_{s}=2g_{s}.  \label{Evsg}
\end{equation}

\subsection{Two Dimensional Dilaton Gravity\label{2D}}

In this section we review some basic facts about 2D gravity which will be
useful in the analysis of the action (\ref{2Dflat}).

Two--dimensional dilaton gravity is a natural way to define, in two
dimensions, theories with a gravitational sector. The basic fields are the
matter fields, as well as the metric $g_{ab}$ and a dilaton field $\Phi $,
which should be considered together as the gravitational fields.
The action has the general form 
\begin{equation*}
S_{\mathrm{2D}}\left( g,\Phi \right) +S_{\mathrm{M}}\left( g,\Phi ,\mathrm{%
Matter}\right) .
\end{equation*}
$S_{\mathrm{M}}$ is the matter part of the action, whereas $S_{\mathrm{2D}}$
is the generalization of the Einstein--Hilbert term and is given by 
\begin{equation*}
S_{\mathrm{2D}}=\int d^{2}x\sqrt{g}\left[ \Phi R-V\left( \Phi \right) \right]
\,,
\end{equation*}
where $V\left( \Phi \right) $ is a potential for the dilaton. The equations
of motion are easily derived to be 
\begin{eqnarray}
2\nabla _{a}\nabla _{b}\Phi &=&g_{ab}\left( 2\square \Phi +V\right) -\tau
_{ab}\,\ ,  \label{metric-eq} \\
R &=&\frac{dV}{d\Phi }+\rho \,,  \notag
\end{eqnarray}
where $\tau _{ab}$ and $\rho $ are 
\begin{equation*}
\tau _{ab}=-\frac{2}{\sqrt{g}}\frac{\delta S_{\mathrm{M}}}{\delta g^{ab}}%
\,,\ \ \ \ \ \ \ \ \ \ \ \ \ \ \ \ \ \ \ \ \ \ \ \ \ \ \ \ \ \ \rho =-\frac{1%
}{\sqrt{g}}\frac{\delta S_{\mathrm{M}}}{\delta \Phi }.
\end{equation*}
Moreover, the conservation of the stress--energy tensor $\tau _{ab}$ is
modified by the dilaton current $\rho $ to 
\begin{equation*}
\nabla ^{a}\tau _{ab}+\rho \nabla _{b}\Phi =0.
\end{equation*}

The inherent simplicity of the dilaton gravity model lies in the following
observations \cite{Kunstatter, Noji, 2Dgravity}. Define $J\left( \Phi \right) $ by 
\begin{equation*}
J=\int Vd\Phi 
\end{equation*}
and consider the function 
\begin{equation}
C=\left( \nabla \Phi \right) ^{2}+J\left( \Phi \right) 
\label{Cfunct}\end{equation}
and the vector field 
\begin{equation*}
\kappa ^{a}=\frac{2}{\sqrt{g}}\epsilon ^{ab}\nabla _{b}\Phi \,.
\end{equation*}
Then, whenever $\tau _{ab}=\rho =0$ -- \textit{i.e. }for any \textit{vacuum}
solution to the equations of motion -- the function $C$ is constant and $%
\kappa $ is a Killing vector of the solution. The first fact follows
immediately from the equations of motion which imply
\begin{equation}
\nabla _{a}C=-\tau _{ab}\nabla ^{b}\Phi +\nabla _{a}\Phi \,(\tau
_{bc}g^{bc})\,.  \label{Cvariation}
\end{equation}
The second fact is proved most easily in conformal coordinates $z^{\pm }$,
with metric 
\begin{equation*}
-dz^{+}dz^{-}e^{\Omega }\,.
\end{equation*}
Then we have that 
\begin{equation*}
\kappa _{\pm }=\mp \nabla _{\pm }\Phi 
\end{equation*}
and the non--trivial Killing equations become $\nabla _{+}\nabla _{+}\Phi
=\nabla _{-}\nabla _{-}\Phi =0$, which now follow from (\ref{metric-eq})
whenever $\tau _{ab}=0$. Finally note that these equations are equivalent to 
\begin{equation*}
\partial _{-}\kappa ^{+}=\partial _{+}\kappa ^{-}=0\,.
\end{equation*}

Using these facts it is trivial to find all classical vacuum solutions. In
fact, whenever $\kappa ^{2}<0$, the metric can be put locally into the form 
\begin{equation}
ds^{2}=-f\left( x\right) dt^{2}+dx^{2}\,,\,\ \ \ \ \ \ \ \ \ \ \ \ \ \ \ \ \
\ \ \ \ \ \ \ \ \ \ \ \ \ \ \ \ \ \kappa =\alpha \frac{\partial }{\partial t}%
\,,  \label{sol1}
\end{equation}
where $\alpha $ is a constant. Then, one has the explicit solution of the
equations of motion 
\begin{equation}
C=\left( \frac{d\Phi }{dx}\right) ^{2}+J\left( \Phi \right) \,,\ \ \ \ \ \ \
\ \ \ \ \ \ \ \ \ \ \ \ \ \ \ \ \ \ \ \ \ \ f=\frac{4}{\alpha ^{2}}\left( 
\frac{d\Phi }{dx}\right) ^{2}\,,  \label{sol2}
\end{equation}
which depends only on the constant $C$. Similar equations hold in regions
where $\kappa ^{2}>0$.

Let us now analyze the geometry in the presence of matter. For our purposes,
we are going to consider only matter lagrangians $S_{M}$ which do not depend
on the dilaton, and which are conformal. This implies that 
\begin{eqnarray*}
&&\tau _{+-} =\rho =0 \,, \\
&&\partial _{-}\tau _{++} = \partial _{+}\tau _{--}=0\,.
\end{eqnarray*}
The simplest example is clearly a conformally coupled scalar $\eta $ with action $%
S_{M}=-\int \left( \nabla \eta \right) ^{2}$. The effect of this type of
matter is best described by considering a shock wave \cite{CGHS,Lawrence},
which is represented in conformal coordinates by a stress--energy tensor of
the form 
\begin{equation*}
\tau _{--}\left( z^{-}\right) =\epsilon \,\delta \left(
z^{-}-z_{0}^{-}\right) \,.\,\ \ \ \ \ \ \ \ \ \ \ \ \ \ \ \ \ \ \ \ \ \ \ \
\left( \epsilon >0\right) 
\end{equation*}
\begin{figure}
\begin{center}
\includegraphics{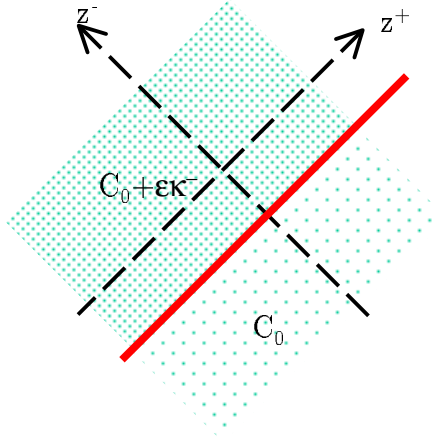}
\end{center}
\caption{Shock wave solution in two--dimensional dilaton gravity.\label{fig6}}
\end{figure}
The positivity of $\epsilon $ can be understood by looking at the
conformally coupled scalar, for which $\tau _{--}=2\left( \nabla _{-}\eta
\right) ^{2}>0$. Recalling from (\ref{Cvariation}) that 
\begin{equation}
\nabla _{-}C=2\tau _{--}\nabla _{+}\Phi \,e^{-\Omega }=\tau _{--}\kappa
^{-}\,,  \label{Cjump}
\end{equation}
we conclude that the shock front interpolates, as we move along $z^{-}$,
between the vacuum solution with $C=C_{0}$ and the vacuum solution with $%
C=C_{0}+\epsilon \kappa ^{-}\left( z_{0}^{-}\right) $ (see Figure \ref{fig6}). As a
consistency check note that, since in the vacuum $\tau _{--}$ and $\kappa
^{-}$ are functions only of $z^{-}$, equation (\ref{Cjump}) defines a jump
in the function $C$ which is independent of the position $z^{+}$ along the
shock wave.

\subsection{The Case $V=2\Phi ^{-3}$}

\begin{figure}
\begin{center}
\includegraphics{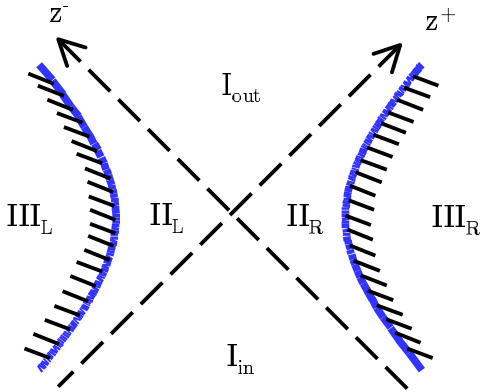}
\end{center}
\caption{The cosmological vacuum solution for $V=2\Phi^{-3}$. Regions I$_{in}$, 
I$_{out}$ are the contracting and expanding cosmologies, whereas regions II$_{L,R}$
are the intermediate regions. Regions III$_{L,R}$ are after the singularities, and 
correspond, when uplifted to three dimensions, to the regions where the
Killing vector $\kappa$ becomes timelike.\label{fig1}}
\end{figure}

We now specialize to the action (\ref{2Dflat}) by taking 
\begin{equation*}
V=\frac{2}{\Phi ^{3}}\,,\ \ \ \ \ \ \ \ \ \ \ \ \ \ \ \ \ \ \ \ \ \ \ \ \ \
\ \ \ \ J=-\frac{1}{\Phi ^{2}}\,.
\end{equation*}
First consider the vacuum solutions. These are easy to obtain by considering
the flat metric (\ref{flat}) written in the coordinates $x^{\pm
}=X^{\pm }e^{\mp EY}$ and $y=\sqrt{E}Y$, where $E>0$ is a constant with
units of energy, 
\begin{equation*}
\frac{\partial }{\partial y}=\sqrt{E}\left( X^{+}\partial _{+}-X^{-}\partial
_{-}\right) +\frac{1}{\sqrt{E}}\,\partial _{Y}
\end{equation*}
is a Killing direction and $x^{\pm }$ are the two--dimensional coordinates.
If one rewrites the metric in the form (\ref{compactification}), it is easy
to check the normalization (\ref{norm}) and to compute the two--dimensional
metric and dilaton field. The explicit computation is done in \cite{CC} and
the result corresponds to the cosmological solution shown in Figure 
\ref{fig1}, where the timelike orientifold singularities
are at a distance of order $E^{-1}$. For future use, we divide the diagram
in various regions I$_{in,out}$, II$_{L,R}$ and III$_{L,R}$, as shown in the
Figure.

Analytic results are simplest in Lorentzian polar coordinates. For example,
in the Rindler wedges II$_{L}$ and II$_{R}$, where $x^{+}x^{-}<0$, one uses
coordinates $t,x$ defined by $x^{\pm }=\pm xe^{\pm t}$ and obtains the static
solution \cite{CC} 
\begin{eqnarray}
ds_{2}^{2} &=&-dt^{2}\left( \frac{x^{2}}{1-E^{2}x^{2}}\right) +dx^{2}\,,
\label{vacuumsol} \\
\Phi  &=&\frac{1}{\sqrt{E}}\sqrt{1-E^{2}x^{2}}\,.  \notag
\end{eqnarray}
It is easy to check that the above solution satisfies (\ref{sol1}) and (\ref
{sol2}) with 
\begin{equation*}
C=-E.
\end{equation*}
As we already discussed in section \ref{orientifold}, these $C<0$ solutions
correspond to non BPS $O8$--$\overline{O8}$ geometries when uplifted to
string theory with equation (\ref{uplift}). Notice that, given 
$E$\thinspace (or equivalently $g_{s}$ by equation (\ref{Evsg}%
)), one has no freedom in the solution, which is unique. Therefore, \textit{%
no fine tuning is required in the initial conditions for the metric and the
dilaton in order to obtain a solution with a bounce and with past and
future cosmological horizons.} Recall also that the conformal field theory description
of the $O\overline O$ system has two distinct moduli, which are the string coupling and
the distance between the orientifolds in string units. We therefore see that
\textit{the backreaction of the geometry ties the two moduli with equation (\ref{Evsg})
and reduces them to a single free parameter.}

We also notice that the $C=0$ solution is clearly singled out and corresponds to 
the supergravity solution of a single BPS $O8$--plane.
The $C>0$ solutions do not have a clear interpretation in string
theory, and it would be nice to understand this point further. 

Before considering more general solutions in the presence of matter, let us
discuss the coordinate change from $x^{\pm }$ to conformal coordinates $%
z^{\pm }$. For the sake of simplicity, and since this is all we will need,
we write formulae for the case $E=1$. We consider then the change of
coordinates 
\begin{equation}
z^{+}z^{-}=e^{2\Phi }\,\frac{\Phi -1}{\Phi +1},\,\ \ \ \ \ \ \ \ \ \ \ \ \ \
\ \ \ \ \ \ \ \ \ \ \frac{z^{+}}{z^{-}}=\frac{x^{+}}{x^{-}},
\label{confcoord}
\end{equation}
where 
\begin{equation*}
\Phi =\sqrt{x^{+}x^{-}+1}.
\end{equation*}
Then, it is simple to show that the metric becomes conformal 
\begin{equation}
ds_{2}^{2}=-\frac{dz^{+}dz^{-}}{z^{+}z^{-}}\left( \frac{\Phi ^{2}-1}{\Phi
^{2}}\right) .
\label{confmetric}\end{equation}

\begin{figure}
\begin{center}
\includegraphics{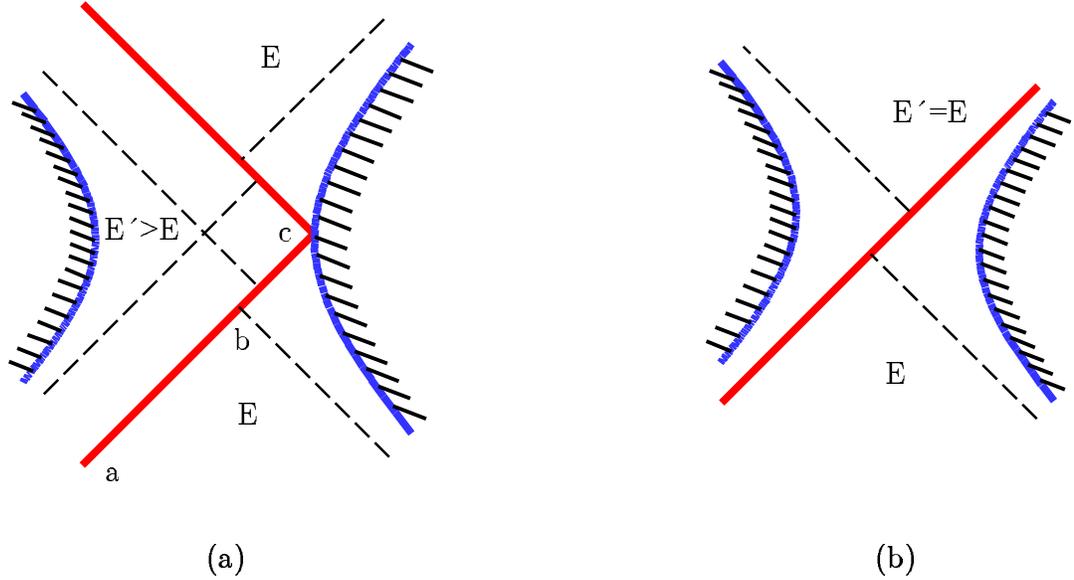}
\end{center}
\caption{Shock wave solutions in the cosmological geometry.\label{fig3}}
\end{figure}

Finally let us add matter and consider shock wave solutions. Given the
discussion in section (\ref{2D}), it is now almost trivial to find the
correct geometries, which are given pictorially in Figure \ref{fig3}. In Figure
\ref{fig3}a, before the shock wave, we have the vacuum solution with some given
value of $E$. After the wave, one has again a vacuum solution, but with a
different value $E^{\prime }$. Recall that 
\begin{equation*}
E^{\prime }=E-\epsilon \kappa ^{-}
\end{equation*}
where $\epsilon >$ $0$ and $\kappa ^{-}=2\nabla _{+}\Phi \,e^{-\Omega }$
must be computed along the wave. As we move in Figure \ref{fig3}a from point $a$ to
points $b$ and $c$ along the shock wave, in the direction of increasing $%
z^{+}$, the value of $\Phi $ decreases to $0$ at $c$ on the singularity.
Therefore $\nabla _{+}\Phi \,<0$ and one has that 
\begin{equation*}
E^{\prime }>E.
\end{equation*}
Moreover, in any vacuum solution with $C=-E$, the value of the
dilaton on the horizons is $1/\sqrt{E}$, as can
be seen from equation (\ref{vacuumsol}) at $x=0$.
Therefore, since the value of $\Phi $ is continuous across the shock wave,
we notice that the horizon to the left of the wave, where the dilaton has
value $\Phi =1/\sqrt{E^{\prime }}$, must intersect the wave between\ the
points $b$ and $c$, as drawn.

A particular case of the shock wave geometry is attained when the wave moves
along one of the horizons, as shown in Figure \ref{fig3}b. Since $\Phi $ is constant
along the horizons, we have that $\nabla _{+}\Phi =0$ and therefore that 
\begin{equation*}
E^{\prime }=E.
\end{equation*}
Let us breifly explain why the horizon at a constant value of $z^+$ shifts 
as one passes the shock wave. It is easy to see that the horizon in question is given by
the curve $\kappa^+ = 0$. In the vacuum, $\kappa^+$ is a function of $z^+$ alone,
but in the presence of matter one has that
\begin{equation*}
\partial_- \kappa^+ = e^{-\Omega} \tau_{--}\,.
\end{equation*}
Then, since $\Omega$ is constant along the horizons (and therefore along the shock
wave) and since $\tau_{--}$ has a delta singularity, the function $\kappa^+$
just jumps by a finite constant across the wave, thus explaining the shift in
the position of the horizon.

Let us conclude by noticing that the instability of the null orbifold found in \cite{Lawrence}
is again a shock wave which interpolates between two different vacuum solutions. 
The initial solution is a BPS vacuum ($C=0$), which has a different spacetime structure
from the other solutions with $C\neq0$ which are formed after the shock wave.
In our case this instability does not arise since we are interpolating between
two non--BPS vacua with the same global structure.

\subsection{The Shock Wave Solution in Three Dimensions}

\begin{figure}
\begin{center}
\includegraphics{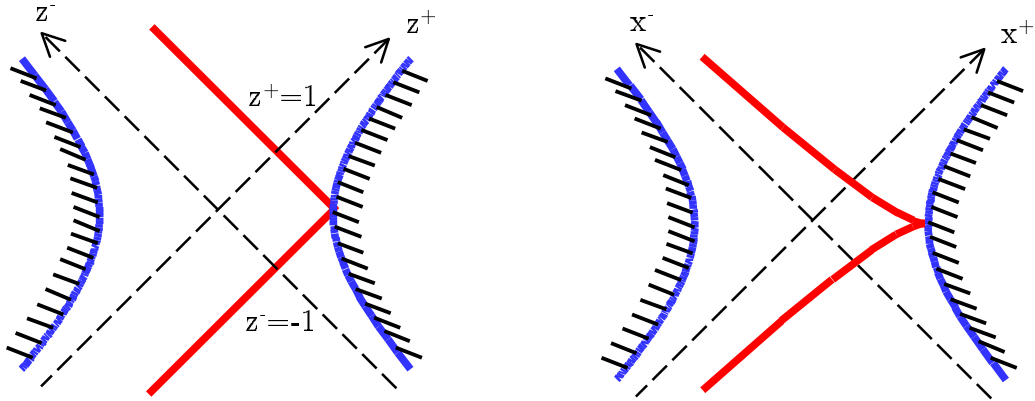}
\end{center}
\caption{The shock wave before backreaction, both in conformal coordinates
$z^\pm$ and in the coordinates $x^\pm$ of equation (\ref{confcoord}).\label{fig2}}
\end{figure}

The geometry of Figure \ref{fig3}a corresponds, when uplifted to three dimensions,
to the gravitational backreaction to matter distributed along a specific
surface, which is the uplift of the shock wave curve. In order to better
understand the geometry of this surface, consider first the shock wave
without backreaction. From now on we take, for simplicity, $E=1$ and $%
E^{\prime }=1+\delta $ for some $\delta >0$. The limiting geometry for $%
\delta \rightarrow 0$ is given by Figure \ref{fig2}. The underlying 2D geometry is
the vacuum $E=1$ solution, and we have singled out a specific curve, which
is given, in the conformal coordinates $z^{\pm }$, by the two segments 
\begin{eqnarray}
z^{-} &=&-1\,,\ \ \ \ \ \ \ \ \ \ \ \ \ \ \ \ \ \ \ \ \ (-\infty <z^{+}<1)
\label{cur} \\
z^{+} &=&1\,.\ \ \ \ \ \ \ \ \ \ \ \ \ \ \ \ \ \ \ \ \ \ \ \
(-1<z^{-}<\infty )  \notag
\end{eqnarray}

We now uplift this configuration to three dimensions. The vacuum solution
becomes, as we emphasized before, nothing but flat space. The curve
becomes then a specific surface in $\mathbb{M}^{3}$, whose geometry is most
easily analyzed by passing to the coordinates $x^{\pm }$ in (\ref{confcoord}%
). It is simple to see, using the explicit coordinate transformation (\ref
{confcoord}), that the curve (\ref{cur}) becomes (see Figure \ref{fig2}) 
\begin{equation}
x^{\pm }=\left( s\pm 1\right) e^{\mp s}  \label{cur2}
\end{equation}
for $s\in \mathbb{R}$. The first part of the curve (\ref{cur}) corresponds
to (\ref{cur2}) for $s<0$, whereas the second part corresponds to $s>0$.
Comparing with (\ref{surxpxm}) we immediately see that this curve
corresponds, when uplifted to three dimensions, to the Horowitz--Polchinski
surface $S_{1}$. Recall from section \ref{HPproblem} that, among the various
surfaces $S_{a}$, the surface $S_{1}$ is special since it is the only null
surface, and it is therefore not surprising that we are obtaining exactly $%
S_{1}$ from the two--dimensional shock wave.

We therefore conclude that \textit{the geometry of Figure \ref{fig3}a, when uplifted
to three dimensions, represents the gravitational backreaction to light rays
distributed on the HP surface }$S_{1}$\textit{.}

\subsection{Finding the Closed Timelike Curve}

We are now in a position to investigate the existence of closed timelike
curves in the geometry of Figure \ref{fig3}a. Recall that, without shock wave,
the quotient geometry has closed timelike curves that always pass
through region III, and that are removed by excising this region. These
curves are \textit{not closed in the covering space}, which is flat. In the
geometry with the shock wave, we shall see that there are closed timelike
curves \textit{already in the covering space}, which could signal an instability. However
these curves always pass through region III, and again should be excised.

In order to describe the closed timelike curves, we first need to prove a simple fact.
Consider, in $\mathbb{M}^{3}$, two points $A$ and $B$ which are connected by
a future--directed timelike curve from $A$ to $B$. Let $x_{A}^{\pm },0$ and $%
x_{B}^{\pm },y_{B}$ be the coordinates of these two points in the coordinate
system (\ref{2dcoord}) and let us assume that both points are in Region II$%
_{R}$, \textit{i.e.} the region with $x^{+}>0$, $x^{-}<0$ and $x^{+}x^{-}<1
$. We parametrize the path from $A$ to $B$ as $x^{\pm }\left( s\right)
,y\left( s\right) $, with $0\leq s\leq 1$. First of all, it is easy to show
that the full path must be entirely regions II$_{R}$ or III$_{R}$%
, because otherwise the horizons in the geometry prevent the curve from
returning to region II$_{R}$ without becoming spacelike. Therefore, since we
are confined to the right Rindler wedge of the $x^{\pm }$ plane, we may
adopt polar coordinates $x^{\pm }=\pm x\,e^{\pm t}$. We want to show that, 
\textit{if }$t_{A}>t_{B}$\textit{\ (the point }$A$\textit{\ is after the
point }$B$\textit{\ in Rindler time), then any forward directed timelike
curve from }$A$\textit{\ to }$B$\textit{\ must go into region III}$_{R}$ 
\textit{(the region with }$x>1$\textit{). }To prove this fact, recall first
the metric on $\mathbb{M}^{3}$ in the coordinates $x,t,y$%
\begin{equation*}
dx^{2}-x^{2}dt^{2}+dy^{2}\left( 1-x^{2}\right) -2x^{2}dydt\,.
\end{equation*}
Consider then the function $t\left( s\right) $. It starts at $t\left(
0\right) =t_{A}$ and it is increasing for small values of $s$. This is
because surfaces of constant $t$ are spacelike in region II$_{R}$ (since $%
dx^{2}+dy^{2}\left( 1-x^{2}\right) >0$ if $x^{2}<1$) and therefore future
directed timelike curves have always $\frac{dt}{ds}>0$ in region II$_{R}$.
Now, since by assumption $t_{B}<t_{A}$, the function $t\left( s\right) $ must
achieve a maximum for some value of $s$. Then, using the same reasoning, the
curve must be in region III$_{R}$ to be timelike, and this concludes the
proof. We can actually always construct a future directed timelike curve by
connecting the two points $A$ and $B$ with a straight line, provided that we
choose $y_{B}$ large enough, as can be seen by direct computation.

We can now prove the existence of a closed timelike curve in the full geometry
(\ref{fig3}a). The regions to the left and to the right of the wave are vacuum 
solutions which correspond, when uplifted to three dimensions, to regions in 
flat Minkowski space. We can then think of the full geometry as follows. Let 
$\widetilde{\mathbb{M}}^{3}$ and $\mathbb{M}^{3}$ be two copies of three--dimensional
space which are going to describe the geometry to the left and to the right
of the shock wave, and which we parametrize with coordinates
\begin{eqnarray*}
&&\widetilde{X}^\pm=\widetilde{x}^{\pm }e^{\pm \sqrt{E^{\prime }}\widetilde{y}}\,\ \ \ \ \ \ \
\ \
\widetilde{Y}=\frac{1}{\sqrt{E^{\prime }}}\ \widetilde{y}\ \ \ \ \ \ \ \ \ \ \ \ \ \ \ \ \
\ \ \ \ \ \ (\widetilde{\mathbb{M}}^{3}) \\
&&X^\pm=x^{\pm }e^{y}\ \ \ \ \ \ \ \ \ \ \ \ \ \ \ \ Y=y\,\ \ \ \ \ \ \ \ \ \ \ \ \
\ \ \ \ \ \ \ \ \ \ \ \ \ \ \ \ \ (\mathbb{M}^{3})\
\end{eqnarray*}
respectively. The shock wave itself defines
surfaces $\widetilde{S}$ and $S$ in the two spaces $\widetilde{\mathbb{M}}^{3}$
and $\mathbb{M}^{3}$. The full geometry is then obtained as follows. Denote
the region to the left of $\widetilde{S}$ in $\widetilde{\mathbb{M}}^{3}$ 
by $\widetilde{\mathbb{M}}^{3}_L$. This corresponds to the part of the 
geometry to the left of the shock wave. Similarly, denote 
the region to the right of ${S}$ in ${\mathbb{M}}^{3}$ by ${\mathbb{M}}^{3}_R$,
which is the region to the right of the wave. The full geometry is then 
given by taking the two regions $\widetilde{\mathbb{M}}^{3}_L$ and
${\mathbb{M}}^{3}_R$ and gluing the two boundaries $\widetilde{S}$
and ${S}$ in a specific way. We do not need the details of the gluing, aside
from the simple fact that $\widetilde{y} = y$ when connecting $\widetilde{S}$
and ${S}$. 
\begin{figure}
\begin{center}
\includegraphics{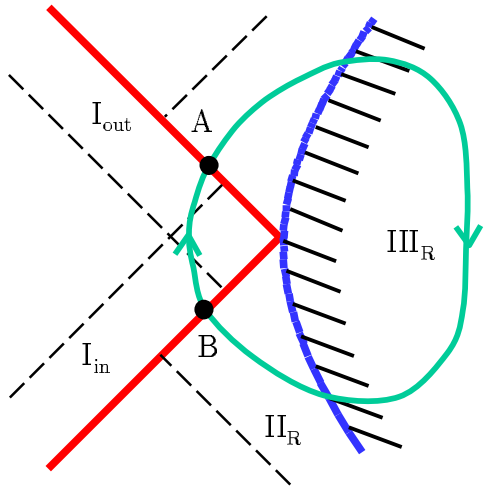}
\end{center}
\caption{A closed timelike curve in the geometry induced by the surface $S_1$.\label{fig4}}
\end{figure}
Looking at Figure \ref{fig4}, which represents the shock wave solution 
in conformal coordinates, choose then the two points $A$ and $B$ as drawn.
They will correspond to points on the surface $\widetilde{S}$ with coordinates 
$\tilde{x}_{A}^{\pm },\tilde{y}_{A}$ and $\tilde{x}_{B}^{\pm },\tilde{y}_{B}$ and
on $S$ with coordinates ${x}_{A}^{\pm },y_A$ and ${x}_{B}^{\pm },{y}_{B}$,
where, as we already noticed, one has $\tilde{y}_{A}={y}_{A}$ and $\tilde{y}_{B}={y}_{B}$.
As in the discussion in the previous part of this section, we choose $y_{A}=0$ and $%
y_{B}$ large and positive.
Notice first that, with respect to the geometry ${\mathbb{M}}^{3}_R$ on the right of the
shock wave, both points $A$ and $B$ are in region II$_{R}$, whereas they are in 
regions I$_{out}$ and I$_{in}$ relative to the geometry $\widetilde{\mathbb{M}}^{3}_L$ 
on the left of the wave. In equations, we have that
\begin{eqnarray*}
x_{A}^{+} &>&0,\,\ \ \ \ \ \ \ \ \ x_{A}^{-}<0,\,\ \ \ \ \ \ \ \ \
x_{A}^{+}x_{A}^{-}<1, \\
x_{B}^{+} &>&0,\,\ \ \ \ \ \ \ \ \ x_{B}^{-}<0,\,\ \ \ \ \ \ \ \ \
x_{B}^{+}x_{B}^{-}<1,\,
\end{eqnarray*}
and
\begin{eqnarray*}
\widetilde{x}_{A}^{\pm } &>&0, \\
\widetilde{x}_{B}^{\pm } &<&0.
\end{eqnarray*}
This is the key feature which is required in order to have a closed timelike curve. 
Let us first concentrate on the right part of the wave. We may just consider the 
straight line from $A$ to $B$ in ${\mathbb{M}}^{3}_R$ which is
future directed for $y_{B}$ large, and which goes, when projected onto the $%
z^{\pm }$ plane as in Figure \ref{fig4}, in the region after the singularity. This
is because the point $A$ is after point $B$ in the Rindler time of region II$%
_{R}$ ($x^+_B/x^-_B>x^+_A/x^-_A$), and so we can apply the arguments of the first part of the section.
Now we consider the region to the left of the shock wave. It is easy to
check, by direct computation, that the straight segment from $B$ to $A$ in 
$\widetilde{\mathbb{M}}^{3}_L$ is still timelike for large values of 
$\tilde{y}_{B}=y_B$, and it is clearly future directed since $B$ is in 
I$_{in}$ and $A$ is in I$_{out}$. This closes the
loop, as drawn in Figure \ref{fig4}, and we have found a closed timelike curve.

Note that, due to the simple lemma proved above, it must be that our closed
curve always goes into region III. Recall though that the boundary of
region III corresponds in string theory to the orientifold $O8$--plane, and
therefore region III should be excised, as described in section \ref
{orientifold}. \textit{The final }$M$\textit{--theory geometry is free of
closed timelike curves, and we have then shown that the presence of the
orientifolds not only resolves the issues put forward in section \ref
{orientifold}, but also cures the instability due to the formation of large
black holes.}

\subsection{Discussion}

Let us conclude this section with some comments.

\begin{figure}
\begin{center}
\includegraphics{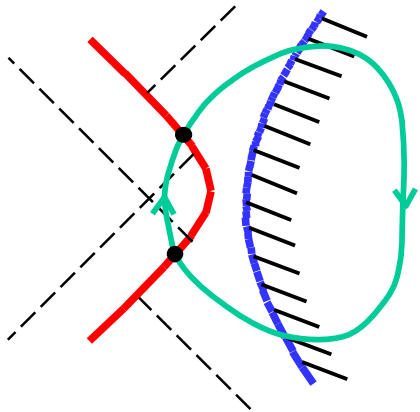}
\end{center}
\caption{A closed timelike curve in the geometry induced by the surface $S_a$
for $a<1$.\label{fig5}}
\end{figure}

First of all, we can try to consider the geometry induced by light rays on
surfaces $S_{a}$ for $-1<a<1$ (so that the surface does not go into region
III). This corresponds, in 2D gravity, to matter which couples to the
dilaton and to the conformal factor of the metric, and is therefore
analytically more complex. Nonetheless, we can understand pictorially, in
Figure \ref{fig5}, that the physics is qualitatively unaltered. In particular
one can find, as before, closed timelike curves which necessarily have to
pass in region III.

Secondly, it is clear that, in Figure \ref{fig3}a, it is crucial that $%
E^{\prime }>E$. This fact, recall, comes from positivity requirements on the
stress--energy tensor $\tau _{ab}$. If $E^{\prime }$ had been less then $E$,
the graph would have had a different structure. In particular, the horizon
of the left part of the geometry would have intersected the shock wave not
between the points $b$ and $c$, but between $a$ and $b$. The causal
structure of the space, when uplifted to three dimensions, is then quite
different and it is easy to show, using arguments similar to those in the
previous section, that the 3D geometry has no closed timelike curves.

Finally, let us comment on the general Horowitz--Polchinski problem for $D>3$.
In this case, gravity is non--trivial even in the absence of matter, and therefore
an exact solution to the problem is probably out of reach. On the other
hand, it seems unlikely that a correct guess on the final qualitative features of the 
solution can be obtained by looking at the interaction between two (or, for that matter,
a finite number) of light--rays. This fact is already true if we just consider the 
\textit{linear reaction} of the gravitational field to the matter distributed
on the HP surfaces $S_a$. Then, very much like in electromagnetism, it is incorrect 
to guess the qualitative features of fields by looking at just a finite subset
of the charges (matter in this case), whenever the charge distribution is infinite
(this infinity is really not an approximation in this case, since it comes
from the infinite extent of the surface $S_a$ due to the unwrapping of the 
quotient space). Therefore, to decide if the problem exists in higher dimensions, 
much more work is required, already in the linear regime of gravity, but most
importantly in the full non--linear setting. Note that \textit{the only case
in which the HP argument is fully correct is exactly in dimension $D=3$, where
the gravitational interaction is topological and when, therefore, the interaction
of an infinite number of charges can be consistently analyzed by breaking it down into 
finite subsets}. This indeed is what we find in the previous sections, since 
closed timelike curves do appear.

\section{Stability of the Cosmological Cauchy Horizon}

Another related classical instability which can arise in orbifold and orientifold
cosmologies is due to the backreaction of general matter fields as they
propagate through the bounce: as the universe contracts, particles will
accelerate and will create a large backreaction in the geometry. General
instabilities of this type can only be studied in the linearized regime, as
opposed to the shock wave geometries which could be analyzed fully,
including gravitational backreaction. In fact, in the previous section we concentrated on
conformally coupled non--dilatonic matter propagating in the
cosmological geometry, and this was possible due to the simple coupling to the
two--dimensional gravitational fields $\Phi $ and $g_{ab}$. We now want to
consider more general matter fields, either non--conformal or
coupled to the dilaton. In the following, we shall concentrate on a scalar
propagating in the quotient space $\mathbb{M}^{3}/e^{\kappa }$, which
corresponds to a free massive scalar on the covering space $\mathbb{M}^{3}$.
Our results will apply to the orientifold cosmology after imposing
Dirichlet or Neumann boundary conditions on the surface $X^{+}X^{-}=-E^{2}$%
. These wave functions were studied in \cite{CCK}, and we shall review
briefly those results before considering the issue of stability.

\subsection{Particle States}

Let us start with a massive field $\Psi $ of mass $m$ on the covering space $%
\mathbb{M}^{3}$, which satisfies the Klein--Gordon equation 
\begin{equation}
\square \,\Psi (X)=m^{2}\,\Psi (X)\ .
\label{KG}\end{equation}
We demand that $\Psi $ be invariant under the orbifold action, so that 
\begin{equation}
\Psi (X)=\Psi (e^{\kappa }X)\ ,  \label{bc}
\end{equation}
where $\kappa =2\pi i\left( \Delta J+RK\right) $ and 
\begin{eqnarray*}
iJ &=&X^{+}\partial _{+}-X^{-}\partial _{-}\ , \\
iK &=&\partial _{Y}\ .
\end{eqnarray*}
The energy scale $E$ of section \ref{mysect} is given by $E=\Delta R^{-1}$.
Since the quotient space has the symmetry generated by the Killing vector $%
\kappa $, and since the operators $J$, $K$ and $\square $ commute, it is
convenient to choose a basis of solutions to (\ref{KG}) and (\ref{bc})
where the operators $J$ and $K$
are diagonal. We then choose our field $\Psi $ so that 
\begin{eqnarray*}
J\,\Psi (X) &=&p\,\Psi (X)\ , \\
K\,\Psi (X) &=&k\,\Psi (X)\ ,
\end{eqnarray*}
by writing 
\begin{equation*}
\Psi _{p,k}\left( X\right) =\Psi _{p}(X^{+},X^{-})\, e^{ikY} \,.
\end{equation*}
The wave functions $\Psi _{p}$ are functions of Lorentzian spin $J=p$ which also
satisfy the two--dimensional Klein--Gordon equation $\left( 4\partial
_{+}\partial _{-}+\omega ^{2}\right) \Psi _{p}=0$, where $\omega
^{2}=m^{2}+k^{2}$. They can then be solved in Lorentzian polar coordinates,
as we shall review below, in terms of Bessel functions, which naturally have
an integral representation as a sum of the standard plane waves in the
covering space. In terms of the quantum numbers $p$, $k$, the orbifold
boundary condition (\ref{bc}) becomes simply 
\begin{equation}
\Delta \,p+\,R\,k=n\in \mathbb{Z\,}.  \label{quant}
\end{equation}
Generally, when $R\neq 0$, the spectrum of $p\in {\mathbb{R}}$ is continuous
for each $n$. On the other hand, for the pure boost orbifold ($R=0$), the
spectrum is discrete since $\Delta \,p=n$. This difference is crucial. \ We
shall see more in detail in the next section that, in the general case, 
\textit{it is possible to define particle states that do not destabilize the
cosmological vacuum on the horizons by properly integrating over the various
spins }$p$, whereas in the pure boost case there is an infinite blue shift
as particles approach the tip of the Minkowski cone \cite{HS}. The situation
is analogous to particles propagating in the ``null boost'' and ``null boost 
$+$ translation'' orbifolds studied in \cite{LMS2}. From now on we consider
the general case $R\neq 0$, and we use as quantum numbers $p$ and $n$, thus
labeling the wave functions $\Psi _{p,n}$. 

To study the behaviour of the wave functions describing the collapse of
matter through the contracting region I$_{in}$ we first move to light--cone coordinates $%
x^{\pm }=X^{\pm }e^{\mp EY}$ and then to polar
coordinates in the Milne wedge of the $x^{\pm }$ plane $x^{\pm }=t\,e^{\pm Ex}$. This means
that, using the quantization condition (\ref{quant}), the wave function $%
\Psi _{p,n}$ is given by 
\begin{equation}
\Psi _{p,n}\left( X\right) =\Psi _{p}^{\left( \pm \right) }(x^{+},x^{-})
\, e^{i\frac{n}{R}Y} \,,  \label{basicsol}
\end{equation}
where the functions $\Psi _{p}^{\left( \pm \right) }$ have the form 
\begin{equation*}
\Psi _{p}^{\left( \pm \right) }=J_{\pm ip}(\omega |t|)\,e^{ipEx}
\end{equation*}
and where $J_{\pm ip}$ are the Bessel functions of imaginary order $%
\pm ip$. In terms of the quantum numbers $p$ and $n$, the frequency $\omega $
satisfies the mass shell condition 
\begin{equation*}
\omega ^{2}=m^{2}+\left( Ep-\frac{n}{R}\right) ^{2}\ .
\end{equation*}

\subsection{Near Horizon Behaviour of the Wave Functions}

We are interested in the limit $t\rightarrow 0$ near the cosmological
horizon, so it is convenient to recall the basic functional form of the
Bessel functions 
\begin{equation*}
J_{\pm ip}(\omega |t|)=\left( \frac{\omega |t|}{2}\right) ^{\pm ip}F_{\pm
ip}\left( \omega ^{2}t^{2}\right) \ ,
\end{equation*}
where $F_{\pm ip}$ is an entire function on the complex plane, which has
the expansion 
\begin{equation*}
F_{\pm ip}(z)=\sum_{k=0}^{\infty }\frac{(-)^{k}}{4^{k}\,k!\,\Gamma (k+1\pm
ip)}\,z^{k}\ .
\end{equation*}
Then, in terms of the two--dimensional light--cone coordinates $x^{\pm }$,
which are well defined throughout the whole geometry, we have 
\begin{equation*}
\Psi _{p}^{\left( \pm \right) }=\left( \frac{\omega |x^{\pm }|}{2}\right)
^{\pm ip}F_{\pm ip}\left( \omega ^{2}x^{+}x^{-}\right) \ .
\end{equation*}
These wave functions are well behaved everywhere except at the horizons $%
x^{\pm }=0$, where there is an infinite blue--shift of the frequency. To see
this, consider the leading behaviour of the wave function $\Psi _{p}^{(+)}$
at both horizons ($x^{+}=0$ or $x^{-}=0$) 
\begin{equation*}
\Psi _{p}^{(+)}\sim \left( \frac{\omega |x^{+}|}{2}\right) ^{ip}\frac{1}{%
\Gamma (1+ip)}\ .
\end{equation*}
Near the horizon $x^{-}=0$ the wave function is well behaved and can be
trivially continued through the horizon. Near $x^{+}=0$, on the other hand,
the wave function has a singularity which can be problematic. In fact, close
to the horizon, the derivative of the field $\partial _{+}\Psi
_{p}^{(+)}\propto \left( x^{+}\right) ^{ip-1}$ diverges as $%
x^{+}\rightarrow 0$, and this signals an infinity energy density, since the
metric near the horizon has the regular form $ds^{2}\simeq -dx^{+}dx^{-}$. This
fact was noted already in \cite{HS,Q1}. 

A natural way to cure the problem is to consider wave
functions which are given by linear superpositions of the above basic
solutions with different values of $p$. The problem is then to understand if
general perturbations in the far past $t\ll -E^{-1}$ will evolve
into the future and create an infinite energy density on the horizon, thus
destabilizing the geometry. This problem is well known in the physics of
black holes where, generically, Cauchy horizons are unstable to small
perturbations of the geometry \cite{PenroseSimpson,CH}. We will show in the
next section that the problem does not arise in this cosmological geometry.
More precisely, we will show that perturbations which are localized in the $x
$ direction in the far past $t\ll -E^{-1}$ (and which are therefore necessarily a
superposition of our basic solutions (\ref{basicsol})) do not induce
infinite energy densities on the horizons of the geometry, and can be
continued smoothly into the other regions II$_{L,R}$ and I$_{out}$ of the
geometry. 

This problem was also considered in \cite{Q2}, where a different conclusion
was reached. We will discuss, at the end of section \ref{last}, the argument of 
\cite{Q2} and describe why it does not apply to physically relevant bounded
perturbations.

\subsection{Analysis of Infalling Matter}

For simplicity we will set, from now on, $E=1$ and $m=0$ and we will focus
uniquely on the uncharged sector $n=0$, which corresponds to pure
dimensional reduction. The three--dimensional action  
for the massless scalar $\Psi$ reduces, in two dimensions, to 
\begin{equation*}
\int d^{2}x\sqrt{g}\Phi \left( \nabla \Psi \right) ^{2}\,.
\end{equation*}
The scalar field $\Psi $ is therefore conformally coupled, but has a non
trivial coupling to the dilaton, and satisfies the equation of motion 
\begin{equation}
\square \Psi +\nabla \ln\Phi \cdot \nabla \Psi =0\,.  \label{EoM1}
\end{equation}
As in the previous section, 
we analyze this equation in the background described by the vacuum
solution (\ref{vacuumsol}), which is given in regions I$_{in,out}$ by 
\begin{eqnarray}
ds^{2} &=&-dt^{2}+\frac{t^{2}}{1+t^{2}}dx^{2}\,,  \label{vacuumSol} \\
\Phi  &=&\sqrt{1+t^{2}}\,,  \notag
\end{eqnarray}
where $x^{\pm }=te^{\pm x}$. We have seen, in the previous section, that the
solutions to (\ref{EoM1}) are known exactly in terms of Bessel functions,
since (\ref{EoM1}) is equivalent to the equation 
$\left( \square _{\mathrm{Flat}}-J^{2}\right) \Psi =0$,
where $J=-i\frac{\partial }{\partial x}$ is the boost operator in the $%
x^{\pm }$ plane and $\square _{\mathrm{Flat}}$ is the Laplacian with respect
to the \textit{flat metric }$-dt^{2}+t^{2}dx^{2}\,$. A different and more
general approach to the solution of (\ref{EoM1}), which is more in line with
the literature on the stability of Cauchy horizons in black hole geometries,
is to rewrite (\ref{EoM1}) in terms of the field 
\begin{equation*}
\Lambda =\Psi \sqrt{\Phi }\,.
\end{equation*}
A simple computation shows that equation (\ref{EoM1}) becomes 
\begin{equation*}
\square \Lambda +\frac{1}{4\Phi ^{2}}\left[ \left( \nabla \Phi \right)
^{2}-2\Phi \square \Phi \right] \Lambda =0\,.
\end{equation*}
We now use the fact that, in the vacuum solution (\ref{vacuumSol}), 
equations (\ref{metric-eq}) and (\ref{Cfunct}) with $C=-1$ imply that
$\square \Phi =-2\Phi ^{-3}$ and that $\left( \nabla \Phi
\right) ^{2}=-1+\Phi ^{-2}$. Therefore
the equation of motion becomes 
\begin{equation}
\square \Lambda +\frac{5-\Phi ^{2}}{4\Phi ^{4}}\Lambda =0\,.  \label{EoM2}
\end{equation}

\begin{figure}
\begin{center}
\includegraphics{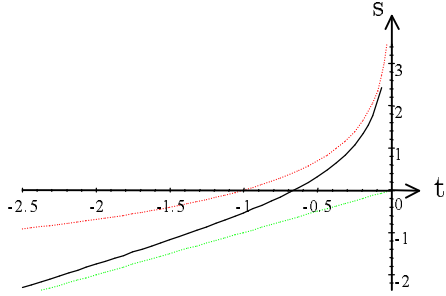}
\end{center}
\caption{The \textit{tortoise} time coordinate $s$ versus $t$, together with the 
asymptotic behaviours $t$ and $-\ln(-t)$ for $t\to -\infty$ and $t\sim 0$.
\label{fig9}}
\end{figure}

As customary, we write the above equation using conformal coordinates
defined in region I$_{in}$. In the sequel, we follow closely the 
beautiful work of Chandrasekar and Hartle \cite{CH}.
We first define, in terms of the coordinates $%
z^{\pm }$ in (\ref{confcoord}), the coordinates $u$, $v$ given by 
\begin{eqnarray*}
z^{+} &=&-e^{-v}\,, \\
z^{-} &=&-e^{-u}\,.
\end{eqnarray*}
The coordinates $u$, $v$ are lightcone coordinates in the $s$, $x$ plane 
\begin{eqnarray*}
u &=&s+x\,, \\
v &=&s-x\,,
\end{eqnarray*}
where $s$ is the usual \textit{tortoise} time coordinate. Recalling (\ref{confcoord}), $s$
is given in terms of $t$ by the expression 
\begin{equation}
s=-\sqrt{1+t^{2}}+\frac{1}{2}\ln \frac{\sqrt{1+t^{2}}+1}{\sqrt{1+t^{2}}-1}\,
\label{tortoise}
\end{equation}
and has asymptotic behaviour for large values of $\left| s\right| $ given by
(see Figure \ref{fig9}) 
\begin{eqnarray*}
s &\sim &t\, ,\ \ \ \ \ \ \ \ \ \ \ \ \ \ \ \ \ \ \ \ \ \ \ \ \ \ \ \ \ \
\ \ \ \ \ \ \ \ \ \ \ \ \left( s\ll -1\right)  \\
s &\sim &-\ln \left( -t\right) \, .\ \ \ \ \ \ \ \ \ \ \ \ \ \ \ \ \ \ \ \
\ \ \ \ \ \ \ \ \ \ \ \left( s\gg 1\right) 
\end{eqnarray*}
The various coordinate systems are recalled in Figure \ref{fig11}a. The metric in
region I$_{in}$ is given, from equation (\ref{confmetric}), by 
$-dudv\,t^{2}\left( 1+t^{2}\right) ^{-1}$, and we conclude
that equation (\ref{EoM2}) reads 
\begin{equation*}
\left( \frac{\partial ^{2}}{\partial s^{2}}-\frac{\partial ^{2}}{\partial
x^{2}}\right) \Lambda \left( s,x\right) =V\left( s\right) \Lambda \left(
s,x\right) \,,
\end{equation*}
where 
\begin{equation}
V=\frac{t^2}{4}\frac{\left(4-t^{2}\right)}{\left(1+t^{2}\right) ^{3}}\, .
\label{potential}
\end{equation}
\begin{figure}
\begin{center}
\includegraphics{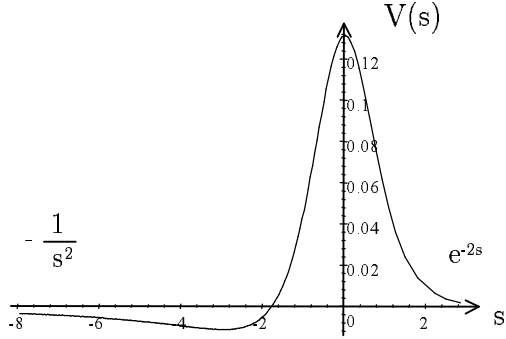}
\end{center}
\caption{The scattering potential $V(s)$.\label{fig10}}
\end{figure} 
Finally, we Fourier transform the $x$ coordinate 
\begin{equation*}
\Lambda \left( s,x\right) =\int dp\,e^{ipx}\,\Lambda \left( s,p\right) 
\end{equation*}
to obtain the Schr\"{o}dinger--like equation 
\begin{equation}
\left( \frac{\partial ^{2}}{\partial s^{2}}+p^{2}\right) \Lambda \left(
s,p\right) =V\left( s\right) \Lambda \left( s,p\right) \,.  \label{ODE}
\end{equation}

Let us consider this differential equation in more detail. The potential $%
V\left( s\right) $, shown in Figure \ref{fig10}, has the following asymptotic
behaviour 
\begin{eqnarray}
V &\sim &e^{-2s}\,\,,\ \ \ \ \ \ \ \ \ \ \ \ \ \ \ \ \ \ \ \ \ \ \ \ \ \ \ \
\ \ \ \ \ \ \ \ \ \ \ \ \ \ \ \ \ \left( s\rightarrow \infty \right) 
\label{asim} \\
V &\sim &-\frac{1}{4s^{2}}\,\ ,\ \ \ \ \ \ \ \ \ \ \ \ \ \ \ \ \ \ \ \ \ \ \
\ \ \ \ \ \ \ \ \ \ \ \ \ \ \ \ \ \left( s\rightarrow -\infty \right)  
\notag
\end{eqnarray}
and is therefore significant only in the region around $s\sim 0$. 
The scattering potential connects the two regions with $s\ll -1$ and $s\gg 1$ (or $%
t\ll -1$ and $t\sim 0$) where the field $\Lambda $ behaves
essentially like a free scalar in two--dimensional flat space. The potential $%
V$ then connects the past Minkowski region ($t\ll -1$ with metric $%
-dt^{2}+dx^{2}$) to the future Milne wedge ($t\sim 0$ with metric $%
-dt^{2}+t^{2}dx^{2}$).
Since $V$ decays at infinity faster then $s^{-1}$, one can follow the usual
theory of one--dimensional scattering. In particular, following \cite{CH}, we
consider the solutions $F\left( s,p\right) $ and $P\left( s,p\right) $ of (%
\ref{ODE}) which behave like pure exponentials $e^{-ips}$ respectively in
the future $s\rightarrow \infty $ and in the past $s\rightarrow -\infty $%
\begin{eqnarray}
F\left( s,p\right)  &\sim &e^{-ips}\sim \left( -t\right) ^{ip}\,,\,\ \ \ \ \
\ \ \ \ \ \ \ \ \ \ \ \ \ \ \ \ \ \ \ \left( s\rightarrow
\infty \right)   \label{as} \\
P\left( s,p\right)  &\sim &e^{-ips}\sim e^{-ipt}\,.\ \ \ \ \ \ \ \ \ \ \ \ \
\ \ \ \ \ \ \ \ \ \ \ \ \ \ \left( s\rightarrow -\infty
\right)   \notag
\end{eqnarray}
Therefore, when restoring the $x$ dependence $e^{ipx}$ of the full solution,
we have, for $s\to\infty$, the plane waves
\begin{eqnarray*}
e^{ipx}F\left( s,p\right)  &\sim &e^{-ipv}\,,  \\
e^{ipx}F\left( s,-p\right)  &\sim &e^{ipu}\,,
\end{eqnarray*}
and, for $s\to-\infty$,
\begin{eqnarray*}
e^{ipx}P\left( s,p\right)  &\sim &e^{-ipv}\,, \\
e^{ipx}P\left( s,-p\right)  &\sim &e^{ipu}\,.
\end{eqnarray*}
As explained in the previous section, the problem is exactly soluble in terms of
Bessel functions\footnote{%
Recall that $J_{\nu }\left( z\right) =z^{\nu }\times \left( \mathrm{%
entire\;function\;of\;}z\right) $. In this paper we choose \textit{%
unconventionally} to take the branch cut of the logarithm along the negative
imaginary axis, so that $z^{\nu }$ and $J_{\nu }\left( z\right) $ also have
a cut on $\func{Re}z=0$, $\func{Im}z<0$. Moreover, the expressions like $%
p^{ip}$ and $\sqrt{p}$ will always be defined using the same prescription
for the logarithm.}, and one has the following explicit form of the
functions $F$ and $P$%
\begin{eqnarray*}
F\left( s,p\right)  &=&\left( \frac{2}{p}\right) ^{ip}\,\Gamma \left(
1+ip\right) \,\,J_{ip}\left( -pt\right) \,\left( 1+t^{2}\right) ^{\frac{1}{4}%
}\,, \\
P\left( s,p\right)  &=&e^{\frac{i\pi }{4}-\frac{\pi p}{2}}\,\sqrt{\frac{\pi p%
}{2}}\,H_{ip}^{\left( 1\right) }\left( -pt\right) \,\left( 1+t^{2}\right) ^{%
\frac{1}{4}}\, ,
\end{eqnarray*}
where $t=t\left( s\right) $ is defined by (\ref{tortoise}). The function
\begin{equation}
H_{ip}^{\left( 1\right) }=\frac{1}{\sinh\left(
\pi p\right)} \left( e^{\pi p}J_{ip}-J_{-ip}\right)
\label{H1}\end{equation}
is the Hankel function of the first kind, which has the correct asymptotic behaviour 
to ensure that $P(s,p)$ has the desired properties.

It is useful,
again following closely \cite{CH}, to consider the general analytic behaviour of $F$
and $P$ as a function of the \textit{complex variable} $p$, at fixed time $s$%
. First it is clear that we must have 
\begin{eqnarray*}
\overline{F\left( s,p\right) } &=&F\left( s,-\overline{p}\right) \,, \\
\overline{P\left( s,p\right) } &=&P\left( s,-\overline{p}\right) \,.
\end{eqnarray*}
\begin{figure}
\begin{center}
\includegraphics{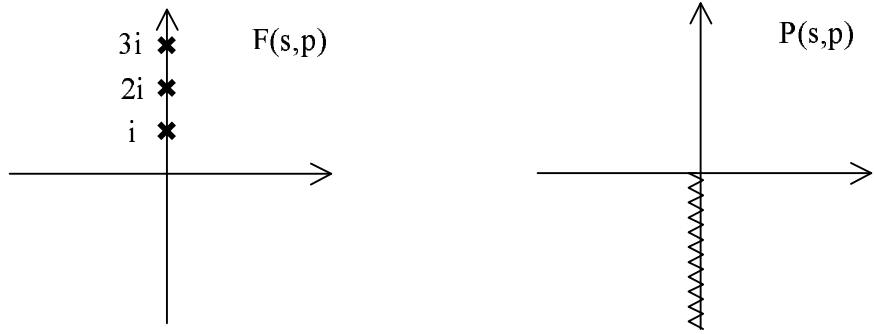}
\end{center}
\caption{The analytic structure of $F(s,p)$ and $P(s,p)$ as functions of $p$.\label{fig12}}
\end{figure}
Moreover, we recall from \cite{CH} that, whenever the potential has asymptotic
behaviour 
\begin{eqnarray*}
V &\sim &e^{-2\kappa _{-}s}\,,\ \ \ \ \ \ \ \ \ \ \ \ \ \ \ \ \ \ \ \ \ \ \
\ \ \ \ \ \ \ \ \ \ \ \ \left( s\rightarrow \infty \right)  \\
V &\sim &e^{2\kappa _{+}s}\ ,\ \ \ \ \ \ \ \ \ \ \ \ \ \ \ \ \ \ \ \ \ \ \ \
\ \ \ \ \ \ \ \ \ \ \ \ \left( s\rightarrow -\infty \right) 
\end{eqnarray*}
then $F\left( s,p\right) $ is everywhere analytic, except for poles at $%
p=i\kappa _{-}\,\mathbb{N}\ $(with $\mathbb{N}$ being the natural numbers $%
1,2,\cdots $), and $P\left( s,p\right) $ has poles at $p=-i\kappa _{+}%
\mathbb{N}$. In our case $\kappa _{-}=1$ and, as expected, the function $F$
has poles at $p=i\mathbb{N}$ due to the gamma function\footnote{%
The function $\left( \frac{2}{p}\right) ^{ip}\,J_{ip}\left( -pt\right) $ is
an entire function of $p$, as can be seen from the explicit power series
representation of the Bessel function $J_{\nu }\left( z\right) $.} $\Gamma
\left( 1+ip\right) $. On the other hand, for the case of the specific
potential (\ref{potential}), one has $\kappa _{+}\rightarrow 0$, with the
exponential behaviour replaced with a power law $s^{-2}$. This is a direct
consequence of the fact that the metric distance from a point in region I$%
_{in}$ to the horizon at $s\rightarrow \infty $ is finite, whereas it is
infinite going to past infinity at $s\rightarrow -\infty $. Therefore, the
poles at $-i\kappa _{+}\mathbb{N}$ become infinitely close and are
effectively replaced by a branch cut along the negative imaginary $p$ axis,
as one can check from the analytic expression for $P$. The features just
described, which are shown in Figure \ref{fig12}, 
are generic and do not depend on the details of the potential,
but only on the asymptotic behaviour (\ref{asim}). 
In particular, one has the same behaviour for fluctuations
of \textit{any field}.

Let us now consider a general solution $\Lambda \left( s,p\right) $ of (\ref
{ODE}). Since both $F\left( s,\pm p\right) $ and $P\left( s,\pm p\right) $
are bases of the solutions of (\ref{ODE}), one must have that 
\begin{eqnarray}
\Lambda \left( s,p\right)  &=&V_{F}\left( p\right) F\left( s,p\right)
+U_{F}\left( p\right) F\left( s,-p\right)   \notag \\
&=&V_{P}\left( p\right) P\left( s,p\right) +U_{P}\left( p\right) P\left(
s,-p\right) \,.  \label{past}
\end{eqnarray}
The coefficients $V_{F,P}$ and $U_{F,P}$ can be easily computed by recalling
that, given two solutions $f$, $g$ of (\ref{ODE}), the Wronskian 
\begin{equation*}
\left[ f,g\right] =f\dot{g}-g\dot{f}
\end{equation*}
is independent of $s$, where the dot denotes differentiation with respect to 
$s$.
\begin{figure}
\begin{center}
\includegraphics{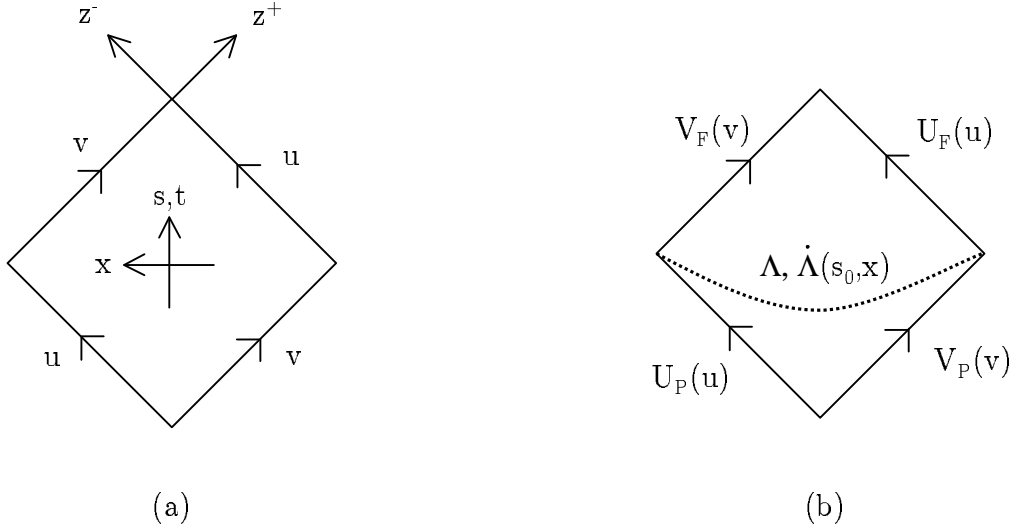}
\end{center}
\caption{Figure (a) shows the various coordinate systems in region I$_{in}$.
Figure (b) shows the various ways
to define a solution $\Lambda$. One either gives the functions $V_P$, $U_P$ or the
functions $V_F$, $U_F$, or, alternatively, the functions $\Lambda(s_0,x)$, 
$\dot{\Lambda}(s_0,x)$ at some fixed time $s_0$ along the dotted line.\label{fig11}}
\end{figure}
In particular, one can easily show that 
\begin{equation*}
\left[ F\left( s,p\right) ,F\left( s,-p\right) \right] =\left[ P\left(
s,p\right) ,P\left( s,-p\right) \right] =2ip\,.
\end{equation*}
Therefore we readily conclude that 
\begin{eqnarray}
V_{F}\left( p\right)  &=&\frac{1}{2ip}\left[ \Lambda \left( s,p\right)
,F\left( s,-p\right) \right] \,,  \label{future} \\
U_{F}\left( p\right)  &=&-\frac{1}{2ip}\left[ \Lambda \left( s,p\right)
,F\left( s,p\right) \right] \,,  \notag
\end{eqnarray}
together with similar equations for $V_{P}$ and $U_{P}$.

\subsection{Behaviour at the Horizons\label{last}}

The meaning of the
coefficients $V_{F}$ and $U_{F}$ is quite clearly understood by considering
the value of the field $\Lambda $ on the horizons. Since we are in the limit 
$s\rightarrow \infty $, we use the asymptotic form of $F\left( s,p\right) $
to conclude that (see Figure \ref{fig11}b)
\begin{equation}
\Lambda \left( v,u\right) \rightarrow V_{F}\left( v\right) +U_{F}\left(
u\right) \,,  \label{a1}
\end{equation}
where 
\begin{equation*}
V_{F}\left( v\right) =\int dp\,V_{F}\left( p\right) e^{-ipv}\,,\ \ \ \ \ \ \
\ \ \ \ \ \ \ \ U_{F}\left( u\right) =\int dp\,U_{F}\left( p\right)
e^{ipu}\,.
\end{equation*}
To compute the stress tensor due to the fluctuation in $\Lambda $, whose
divergence usually signals the instability of the Cauchy horizon, we also
must consider the derivatives of the field with respect to the coordinates $%
z^{\pm }$ (the coordinates $u$, $v$ tend to $\infty $ on the horizons).
Recalling that 
\begin{equation*}
\partial _{+}=-e^{v}\partial _{v}\, ,\ \ \ \ \ \ \ \ \ \ \ \ \ \ \ \ \ \ \ \ \
\ \ \ \ \ \ \ \ \ \ \ \ \ \partial _{-}=-e^{u}\partial _{u}\, ,
\end{equation*}
we conclude that 
\begin{eqnarray}
\partial _{+}\Lambda \left( v,u\right)  &\rightarrow &-e^{v}\partial
_{v}V_{F}\left( v\right) \,,  \label{a2} \\
\partial _{-}\Lambda \left( v,u\right)  &\rightarrow &-e^{u}\partial
_{u}U_{F}\left( u\right) \,.  \notag
\end{eqnarray}
Note that both expressions (\ref{a1}) and (\ref{a2}) are valid \textit{only in
the limit }$( u\rightarrow \infty \mathit{,\ }$ $v\mathit{\ }\mathrm{fixed}%
) $\textit{\ or }$\left( v\rightarrow \infty \mathit{,\ }u\mathit{\ }%
\mathrm{fixed}\right) $\textit{, and are not approximate expressions for
large but finite }$u$\textit{, }$v$\textit{. }This is because, although the
asymptotic form (\ref{as}) is exactly valid in the limit $s\rightarrow
\infty $, it becomes accurate for $s\gg \overline{s}\left( p\right) $, where
the threshold value $\overline{s}\left( p\right) $ depends explicitly on $p$%
. Given the explicit form of $F$, it is clear that $\overline{s}\left(
p\right) \sim \ln p$ since only in this case is the argument of the Bessel
function $-pt\ll 1$. Therefore, given a function $\Lambda $ which has
support at all momenta, one is not uniformly in the asymptotic region for
all momenta at any large but finite $s$. Only in the limit $s\rightarrow
\infty $ we can use uniform convergence to deduce the expressions (\ref{a1})
and (\ref{a2}), in the sense just described.

Finally, going back to the computation of the derivatives $\partial _{\pm }\Lambda $,
one recalls from \cite{CH} that the problematic limits are given by $%
\lim_{v\rightarrow \infty }\,\partial _{+}\Lambda $ and $\lim_{u\rightarrow
\infty }\,\partial _{-}\Lambda $. Concentrating on the first one for
simplicity, we see that we need to consider the expression 
\begin{eqnarray}
\lim_{v\rightarrow \infty }\,\partial _{+}\Lambda  &=&\lim_{v\rightarrow
\infty }\,e^{v}\int dp\,V_{F}\left( p\right) \,ip\,e^{-ipv}  \label{problem}
\\
&=&\frac{1}{2}\lim_{v\rightarrow \infty }\,e^{v}\int dp\left[ \Lambda \left(
s,p\right) ,F\left( s,-p\right) \right] \,e^{-ipv}\,.  \notag
\end{eqnarray}
The physical problem we need to address is the following. Let us
assume that, at some time $s_{0}\ll 0$ in the past, much before the
scattering potential $V\left( s\right) $ becomes relevant, our field $%
\Lambda $ is in some specific configuration given by initial conditions on
the value of $\Lambda $ and of its time derivative $\dot{\Lambda}$ 
\begin{equation*}
\Lambda \left( s_{0},x\right) \,,\ \ \ \ \ \ \ \ \ \ \ \ \ \ \ \ \ \ \ \ \ \
\ \dot{\Lambda}\left( s_{0},x\right) \,.
\end{equation*}
We will assume that the functions above are \textit{localized} as a function
of $x$. More precisely, we shall demand that $\Lambda \left( s_{0},x\right) $
and $\dot{\Lambda}\left( s_{0},x\right) $ are such that their evolution
would not lead to an inconsistent behaviour on the horizon in the Milne wedge
of \textit{flat space}, where $V=0$.
It is important to realize that those are the general physical perturbation, since
we are only demanding that they would be regular in flat space.
Let us be more explicit. In the absence of the potential $V$, the solutions
of the massless Klein--Gordon equations are just $\Lambda = \alpha(v) + \beta(u)$,
and the initial conditions $\Lambda$ and $\dot\Lambda$ at
$s_0$ are equivalent to giving the functions $\alpha$ and $\beta$ of the light--cone
coordinates. Recall though that the coordinates which are regular across the horizon 
are $z^\pm$, and therefore we must demand, in order to have a regular perturbation 
in flat space, that $\alpha$ and $\beta$ be well--behaved \textit{as functions of}
$z^\pm$. To see what this means in practice, consider a function of $z^+$ which
has a nice power series expansion $a_0 + a_1 z^+ + \cdots$ around $z^+=0$.
As a function of $v$ this reads  $\alpha(v) = a_0 - a_1 e^{-v} + \cdots$, and this shows
that, aside from the constant part, the function $\alpha(v)$ must decay, for $v\to\infty$,
at least as fast as $e^{-v}$. This fact imposes constraints on the Fourier transforms
of $\alpha$ and $\beta$ and, in turn, on the functions $\Lambda(s_0,x)$ and $\dot{\Lambda}
(s_0,x)$. In practice, it is sufficient to require that the Fourier transforms $\Lambda
\left( s_{0},p\right) $ and $\dot{\Lambda} \left( s_{0},p\right) $ do not have
poles in the strip $\left| \func{Im}p\right| <1$. Then the problematic limit 
\begin{equation*}
\frac{1}{2}\lim_{v\rightarrow \infty }\,e^{v}\int dp\left[ \Lambda \left(
s_{0},p\right) \dot{F}\left( s_{0},-p\right) -\dot{\Lambda} \left( s_{0},p\right)
F\left( s_{0},-p\right) \right] \,e^{-ipv}
\end{equation*}
can be computed by deforming the contour in the lower half of the complex
plane, and is determined by the poles of $F\left( s,-p\right) $. In fact,
the leading behaviour of the integral is controlled by the first pole at $-i$%
, which gives an asymptotic behaviour for large $v\rightarrow \infty $ of $%
e^{-v}$. The limit in the above expression then tends to a finite result.

\begin{figure}
\begin{center}
\includegraphics{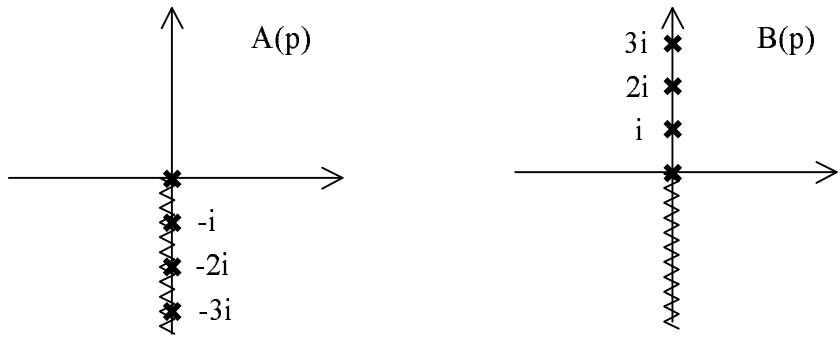}
\end{center}
\caption{The analytic structure of $A(p)$ and $B(p)$.\label{fig13}}
\end{figure}

In order to better understand the above result, let us review the standard
argument for the instability of the Cauchy horizon, which was followed in \cite{Q2}.
The problem is usually posed by giving initial conditions at past null infinity 
(see also Figure \ref{fig11}b), by giving the functions 
\begin{equation*}
V_{P}\left( v\right) =\int dp\, V_{P}\left( p\right) e^{-ipv}\, ,\ \ \ \ \
\ \ \ \ \ \ \ \ \ \ \ U_{P}\left( u\right) =\int dp\, U_{P}\left( p\right)
e^{ipu}\, ,
\end{equation*}
on which we impose a regularity constraint (say, for simplicity, a
localization condition like the one discussed above, now in the coordinates $%
v$ and $u$). To determine $V_{F}$, $U_{F}$ in terms of $V_{P}$, $U_{P}$ one
uses the expression (\ref{past}) for $\Lambda $ in the equation (\ref{future}%
) and obtains that 
\begin{equation*}
\left( 
\begin{array}{c}
V_{F}\left( p\right)  \\ 
U_{F}\left( p\right) 
\end{array}
\right) =\left( 
\begin{array}{cc}
A\left( p\right)  & B\left( -p\right)  \\ 
B\left( p\right)  & A\left( -p\right) 
\end{array}
\right) \left( 
\begin{array}{c}
V_{P}\left( p\right)  \\ 
U_{P}\left( p\right) 
\end{array}
\right) \,,
\end{equation*}
where\footnote{%
The coefficients $A\left( p\right) $ and $B\left( p\right) $ satisfy,
generically, the conjugation relations $\overline{A\left( p\right) }=A\left( -\overline{p}\right)$
and $\overline{B\left(p\right) }=B\left( -\overline{p}\right)$, 
and the unitarity relation 
$A\left( p\right) A\left( -p\right) -B\left( p\right) B\left( -p\right) =1$.
Using (\ref{H1}), we have the explicit
expression 
\begin{eqnarray*}
&& A\left( p\right) = e^{\frac{\pi }{2}p+\frac{i}{4}\pi }\sqrt{\frac{\pi p}{2}}%
\left( \frac{p}{2}\right) ^{ip}\frac{1}{\sinh \left( \pi p\right) }\frac{1}{%
\Gamma \left( 1+ip\right) }\,, \\
&& B\left( p\right) = -e^{-\frac{\pi }{2}p+\frac{i}{4}\pi }\sqrt{\frac{\pi p}{2}}%
\left( \frac{p}{2}\right) ^{-ip}\frac{1}{\sinh \left( \pi p\right) }\frac{1}{%
\Gamma \left( 1-ip\right) }\,.
\end{eqnarray*}
Moreover, in general, the function $A(p)$ will have zeros on the \textit{positive
imaginary $p$--axis} whenever the potential $V$ has a bound state. The explicit expressions
above show that, for our particular potential, there are no bound states.
} 
\begin{eqnarray*}
A\left( p\right)  &=&\frac{1}{2ip}\left[ P\left( s,p\right) ,F\left(
s,-p\right) \right] \,, \\
B\left( p\right)  &=&-\frac{1}{2ip}\left[ P\left( s,p\right) ,F\left(
s,p\right) \right] \,.
\end{eqnarray*}
The above expression shows that both $A\left( p\right) $ and $B\left(
p\right) $ have a cut along the negative imaginary $p$ axis (the limit of
the poles at $-i\kappa _{+}n$ for $\kappa _{+}\rightarrow 0$). The 
full analytic structure of $A(p)$ and $B(p)$ is shown in Figure \ref{fig13}. Since $%
V_{F}\left( p\right) =V_{P}\left( p\right) A\left( p\right) +U_{P}\left(
p\right) B\left( -p\right) $ we conclude that the leading pole of $V_P(p)A(p)$
at $-i\kappa
_{+}\rightarrow -i0$ determines the asymptotic behaviour of the integral in (%
\ref{problem}) to be $e^{-\kappa _{+}v}$. Therefore, the limit $%
\lim_{v\rightarrow \infty }e^{v\left( 1-\kappa _{+}\right) }$ diverges, thus
naively signaling an instability of the horizon.

The above reasoning depends crucially on the assumption that the
functions $U_{P}$, $V_{P}$ have a nice Fourier transform (analytic at least
in the region $\left| \func{Im}p\right| <1$), so that the perturbation is
localized on the null directions at past null infinity. If this is the case,
though, expression (\ref{past}) shows that the function $\Lambda \left(
s,x\right) $ is already delocalized (the Fourier transform has a branch cut
due to $P\left( s,p\right) $) at \textit{any finite time }$s\ll -1$\textit{\
much before the scattering potential}, and therefore will concentrate on the
horizons and create an infinite stress tensor. This is clearly not the type
of perturbation we want to focus on, which should be localized in space
before they hit the potential $V$.
Note that, at any finite time $s$, we cannot use, in equation (\ref{past}),
the asymptotic form of $P(s,p)$ for all values of $p$. In fact, the 
explicit form of the the function $P$ shows that one is in the 
asymptotic regime for $|p|\gg s^{-1}$. Therefore, the crucial low
momentum modes which determine the asymptotic behaviour of the wave
function are \textit{never in the asymptotic region for any
finite time $s$}. We then consider the requirement of
localization of $\Lambda \left( s,x\right) $ at times $s\ll -1$ to be the
correct and physically relevant boundary condition to study issues of
stability of the geometry. The reasonings of this section then show that, in
this sense, the Cauchy horizon in perfectly stable to small perturbations.

\subsection{Discussion}

Let us conclude this section with some comments on future research.

We have shown that, in the orientifold cosmology, one can define particle--like
perturbations which do not destabilize the Cauchy horizon, and therefore that
the solution is stable against small variations of the background fields.
All the work is done at the classical linearized level. The most pressing
question for issues of stability is the study of the quantum stress--energy
tensor for the field $\Psi$. This is a non--trivial problem due to the non--minimal
coupling to the metric and the dilaton, and only some results are known in 
the literature (see \cite{Noji, 2Dgravity} and references therein). On the other
hand, the problem might be tractable since the exact wave--functions are known
in this case.

\section{Conclusion}

In this paper we investigate the classical stability of a two--dimensional orientifold cosmology, 
related to a time--dependent orbifold of flat space--time. We saw, with a specific 
counter example, that the instability argument of Horowitz and Polchinski is not valid once the time--like
orbifold singularity is interpreted as the boundary of space--time. In this specific example, we consider
an exact shock--wave solution of two--dimensional dilaton gravity that uplifts to a distribution of matter 
in the three--dimensional covering space. According to Horowitz and Polchinski this distribution should 
interact gravitationally creating large black holes. Indeed, there are closed time--like curves in the covering space 
geometry of the shock--wave, signaling the three--dimensional gravitational instability.
However, such CTC's are not present if we interpret the singularity as a boundary of space--time, and accordingly 
excise from the geometry the region behind it.

The other stability problem addressed in this work is related to cosmic censorship. The presence of naked 
singularities with a Cauchy Horizon has lead many people to believe that such singularities 
never form. Indeed, as for the Reissner--Nordstrom black hole one expects the Cauchy horizon to be unstable
when crossed by matter. We analyze the propagation of a scalar field coupled to gravity in the geometry, 
and show that any localized fluctuation at some finite time in the far past will not destabilize
the horizon. The existence of the time--like singularity does {\it not} imply the break--down
of predictability because of the conjectured duality between the singularity and orientifolds of 
string theory. To leading approximation, the effect of the orientifolds is to enforce a boundary condition
on the fields, which determines uniquely their evolution. 
A more complete quantum understanding of this system is desirable.

\section*{Acknowledgements}

We would like to thank J. Barbon, C. Herdeiro, F. Quevedo and I. Zavala 
for useful discussions and correspondence. LC is supported by a Marie Curie 
Fellowship under the European Commission's Improving Human Potential programme 
(HPMF-CT-2002-02016). This work was partially supported by CERN under contract 
CERN/FIS/43737/2001.

\end{document}